\documentclass[english]{article}
\usepackage[T1]{fontenc} \usepackage[latin9]{inputenc} 
\usepackage{amsthm}
\PassOptionsToPackage{normalem}{ulem} 
\usepackage{amsmath,amssymb,babel,cite,color,dsfont,empheq,framed,geometry,graphicx,mathrsfs,needspace,setspace,tabularx,tikz,ulem,url}

\usepackage[capbesideposition=inside,
facing=yes,capbesidesep=quad]{floatrow}

\makeatletter

\let\@fnsymbol\@arabic

\makeatother

\newcommand{\Csecond}{\widetilde{\mathbb{C}}}

\newcommand{\Cfourth}{\overline{\mathbb{C}}}
\newcommand{\Csym}{\mathbb{C}}
\newcommand{\Cso}{\Csym}
\newcommand{\Jfourth}{\overline{\Jsym}}
\newcommand{\Jsym}{\mathbb{J}}
\newcommand{\Jso}{\Jsym}
\newcommand{\Lf}{\overline{\mathbb{L}}}
\newcommand{\Ls}{\widehat{\mathbb{L}}}
\newcommand{\Lsym}{\mathbb{L}}
\newcommand{\E}{\overline{\mathbb{E}}_{\mathrm{cross}}}

\newcommand{\Cc}{\Cso_{c}}
\newcommand{\Ctc}{\Csecond_{c}}
\newcommand{\Ce}{\Csym_{e}}
\newcommand{\Cte}{\Csecond_{e}}
\newcommand{\Coe}{\Cfourth_{e}}
\newcommand{\Ch}{\Csym_{\mathrm{micro}}}
\newcommand{\Cth}{\Csecond_{\mathrm{micro}}}
\newcommand{\C}{\Csym_{\mathrm{macro}}}
\newcommand{\Ct}{\Csecond_{\mathrm{macro}}}
\newcommand{\mm}{\mu_{\mathrm{macro}}}
\newcommand{\lm}{\lambda_{\mathrm{macro}}}
\newcommand{\mh}{\mu_{\mathrm{micro}}}
\newcommand{\lh}{\lambda_{\mathrm{micro}}}
\newcommand{\me}{\mu_{e}}
\newcommand{\mc}{\mu_{c}}
\newcommand{\lle}{\lambda_{e}}
\newcommand{\ke}{\kappa_{e}}
\newcommand{\kh}{\kappa_{\mathrm{micro}}}
\newcommand{\km}{\kappa_{\mathrm{macro}}}
\newcommand{\mLc}{\mu L_{c}^{2}}

\newcommand{\R}{\mathbb{R}}
\newcommand{\nablau}{\,\nabla u\,}
\newcommand{\p}{\,{P}}
\newcommand{\Curl}{\,\mathrm{Curl}}
\newcommand{\dev}{\, \mathrm{dev}}
\newcommand{\Div}{\mathrm{Div}}
\newcommand{\tr}{\, \mathrm{tr}}
\newcommand{\sym}{\, \mathrm{sym}}
\newcommand{\Sym}{\mathrm{Sym}}
\renewcommand{\skew}{\, \mathrm{skew}}
\newcommand{\axl}{\,\mathrm{axl}}
\newcommand{\so}{\mathfrak{so}}
\newcommand{\X}{\,}
\newcommand{\x}{\cdot}
\newcommand{\langlenew}{\,\big\langle\,}
\newcommand{\ranglenew}{\,\big\rangle}

\newcommand*\widefbox[1]{\fbox{\hspace{2em}#1\hspace{2em}}}

\title{\vspace{-1.0cm}Transparent anisotropy for the relaxed micromorphic
	model: macroscopic consistency conditions and long wave length asymptotics}

\author{Gabriele Barbagallo\footnote{Gabriele Barbagallo, gabriele.barbagallo@insa-lyon.fr, LaMCoS-CNRS \& LGCIE, INSA-Lyon, Universitité de Lyon, 20 avenue Albert Einstein, 69621, Villeurbanne cedex, France},
	Angela Madeo\footnote{Angela Madeo, corresponding author, angela.madeo@insa-lyon.fr, LGCIE, INSA-Lyon, Université de Lyon, 20 avenue	Albert Einstein, 69621, Villeurbanne cedex, France}
	, 
	Marco Valerio 	d'Agostino\footnote{Marco Valerio d'Agostino, marco-valerio.dagostino@insa-lyon.fr, LGCIE, INSA-Lyon, Université de Lyon, 20 avenue Albert Einstein, 69621, Villeurbanne cedex, France} , 
Rafael Abreu\footnote{Rafael Abreu, abreu@uni-muenster.de, Institut für Geophysik, Westfälische Wilhelms-Universität Münster, Corrensstraße 24, 48149, Münster,  Germany},	\\
	Ionel-Dumitrel Ghiba\footnote{Ionel-Dumitrel Ghiba,  dumitrel.ghiba@uni-due.de, dumitrel.ghiba@uaic.ro,  Chair for Nonlinear Analysis and Modelling, Fakultät für Mathematik, Universität Duisburg-Essen, Thea-Leymann Str.  9, 45127 Essen, Germany; Alexandru Ioan Cuza University of Ia{\c{s}}i, Department of Mathematics, Blvd.  Carol I, no.  11, 700506 Ia{\c{s}}i, Romania; and Octav Mayer Institute of Mathematics of the Romanian Academy, Ia{\c{s}}i Branch, 700505 Ia{\c{s}}i.
}\,
and 
	Patrizio Neff\,\footnote{Patrizio Neff, patrizio.neff@uni-due.de, Head of Chair for Nonlinear Analysis and Modelling, Fakultät für Mathematik, Universität Duisburg-Essen,  Mathematik-Carrée, Thea-Leymann-Straße 9, 45127 Essen}}

\begin{document}
	
\maketitle 
\addtocounter{footnote}{6} 
\vspace{-0.8cm}
\begin{center}
\emph{``Pluralitas non est ponenda sine necessitate.'' - ``Plurality should not to be supposed without necessity.''}\par\end{center}
\begin{center}
John Duns Scotus - ``Ordinatio''\par\end{center}
\vspace{-0.2cm}	
\begin{abstract}
{In this paper, we study the anisotropy classes of the fourth order elastic tensors of the relaxed micromorphic model, also introducing their second order counterpart by using a Voigt-type vector notation. In strong contrast with the usual micromorphic theories, in our relaxed micromorphic model} only classical elasticity-tensors with at most 21 independent components are studied together with rotational coupling tensors with at most 6 independent components. {We show that in the limit case $L_c\rightarrow 0$ (which corresponds to considering very large specimens of a microstructured metamaterial) the meso- and micro-coefficients of the relaxed model can be put in direct relation with the macroscopic stiffness of the medium via a fundamental homogenization formula.}  We also show that a similar {homogenization formula} is not possible {in the case of} the standard Mindlin-Eringen-format of the anisotropic micromorphic model. Our results
  {allow us to forecast the successful short term application of the relaxed micromorphic model to the characterization of anisotropic mechanical metamaterials.}
\end{abstract}

\vspace{0.8cm}

\hspace{-0.55cm}\textbf{Keywords}: relaxed micromorphic model, anisotropy, arithmetic mean, geometric mean, harmonic mean, Reuss-bound, Voigt-bound, generalized continuum models, long wavelength limit, macroscopic consistency, Cauchy continuum, homogenization, multi-scale modeling, parameter identification, non-redundant model 

\vspace{0.7cm}

\hspace{-0.55cm}\textbf{AMS 2010 subject classification}:  74A10 (stress), 74A30 (nonsimple materials), 74A35 (polar materials), 74A60 (micromechanical theories), 74B05 (classical linear elasticity), 74E10 (anisotropy), 74E15 (crystalline structure), 74M25 (micromechanics), 74Q15 (effective constitutive equations)

\newpage

\tableofcontents

\newpage


\section{Prelude}

Modeling in continuum mechanics is an art encompassing mathematics, mechanics, physics and experiments. Many researchers have been attracted to the field of generalized continuum mechanics, following the works of the masters Mindlin and Eringen and have {dealt with} the description of particular aspects of generalized continuum theories,   {usually introducing ``ad hoc'' terms to provide sensational additional effects}. That has been done, while fundamental questions concerning the range of applicability or the descriptive power of generalized continuum mechanics had not been settled leading to an understandable disdain of the majority of researchers in continuum mechanics for these models. We, on the contrary, believe in the usefulness of generalized continuum mechanical models but at the same time we {are aware of} their current shortcomings.

{We are deeply convinced that scientific advancements do} not consist in producing a zoo of possibilities and to combine more effects (which are themselves not yet properly understood), but in \textbf{reducing complexity} and in explaining in simpler terms previously non-connected ideas without losing the accuracy of the mathematical description of the physical problem {we are interested in}.

A major guidance for enlightened modeling certainly comes from the experimental side. Basing ourselves on the phenomena we want to describe, we should not use superfluous information (superfluous because in practice, it cannot be determined) and, among valid competing hypotheses, the one with the \textbf{simplest} assumptions should be selected.

In this work we deal with the anisotropic relaxed micromorphic model in this spirit directed towards simplification. Whether we have achieved a step into this direction must be judged by our readers. 
 
 {
 	This paper originated from the need of setting up a transparent theory for the description of anisotropic materials with embedded microstructures.
 	
 	Recent papers \cite{madeo2016first,madeo2016modeling} provided the evidence that the relaxed micromorphic model, even when restricted to the isotropic case, is usable to characterize the mechanical behavior of metamaterials with unorthodox dynamical properties. More precisely, it has been shown that the isotropic relaxed micromorphic model can be effectively used to model band-gap metamaterials, i.e. microstructured materials which are able to ``stop'' the propagation of elastic waves due to local resonances at the level of the microstructure.
 	
 	The enormous advantage of using the relaxed micromorphic model for the description of such metamaterials is undoubtedly that of mastering the behavior of complex media via the introduction of few elastic coefficients (Young modulus, Poisson ratio and few extra microstructure-related homogenized coefficients). This simplified modeling of metamaterials allows to open the door towards the conception of ``metastructures'', i.e. structures which are made up of metamaterials as basic building blocks and which preserve their unconventional behavior at the scale of the structure (i.e. wave absorption).
 	
 	The successful fitting of the isotropic relaxed micromorphic model on actual metamaterials was already been provided in  \cite{madeo2016first,madeo2016modeling}. Nevertheless, such studies have also suggested the need of generalizing the theoretical framework to the anisotropic case, to describe in detail the mechanical behavior of periodic and quasi-periodic metamaterials.
 	
 	In this paper we provide such theoretical framework with the aim of readily applying it in a forthcoming paper which will be focused on the fitting of the anisotropic relaxed micromorphic model on actual metamaterials with low degree of anisotropy.

 }

\section{Introduction}

Recent years have seen a colossal increase of interest in so called generalized {or enriched} continuum models. This {exponential growth is mainly} due to the need felt to incorporate, viewed from the phenomenological level, additional features like the discreteness of matter, characteristic length scales, dispersion of waves, among others. All such features are not captured by standard elasticity approaches.
{ The idea of using generalized continuum models to account for the homogenized behavior of microstructured materials has extensively been exploited in the last years}  (see e.g \cite{forest1998mechanics,forest2002homogenization,forest2006nonlinear,forest2009micromorphic,forest2011generalized}). 
 One of the most known generalized continuum models is the micromorphic continuum { model} introduced by Mindlin and Eringen \cite{eringen1999microcontinuum,eringen1964nonlinear,eringen1966mechanics,eringen1969micromorphic,claus1971dislocation,mindlin1964micro} in the early sixties of the last century. It includes many special cases among which the much older Cosserat-type models \cite{cosserat1909theorie,neff2006cosserat,neff2009new,jeong2009numerical,jeong2008existence,boehmer2015soliton,lankeit2016integrability,neff2010stable,fischle2015geometrically}.

In this paper we do not {present} the historic development { of enriched continua}, referring the reader to \cite{neff2014unifying,maugin2016nonclassical} for this purpose. Furthermore, we restrict our attention to the linearized framework noting that the first existence result for the geometrically nonlinear static case has been obtained in \cite{jeong2008existence}, which includes a previous result for the nonlinear Cosserat model \cite{neff2004material}. {For more details} about existence results for  micromorphic models at finite deformations, we refer the reader to \cite{neff2015existence,neff2014existence,neff2006existence,lankeit2016integrability}.
 Further existence results are supplied in \cite{mariano2005computational,mariano2009ground,ebobisse2010existence,ebobisse2015existence}. There are many applications treated within the nonlinear micromorphic framework, among which we limit ourselves to mention \cite{kirchner2005unifying,hirschberger2007deformational,sansour1998unified,sansour2010formulation,grammenoudis2009micromorphic,grekova2005modelling,iesan2001extremum,janicke2009two,lazar2007conservation,vernerey2007multi,vernerey2008micromorphic,romano2016micromorphic,meenen2011variationally}.
\medskip

\begin{figure} \fcapside{\includegraphics[width=7cm]{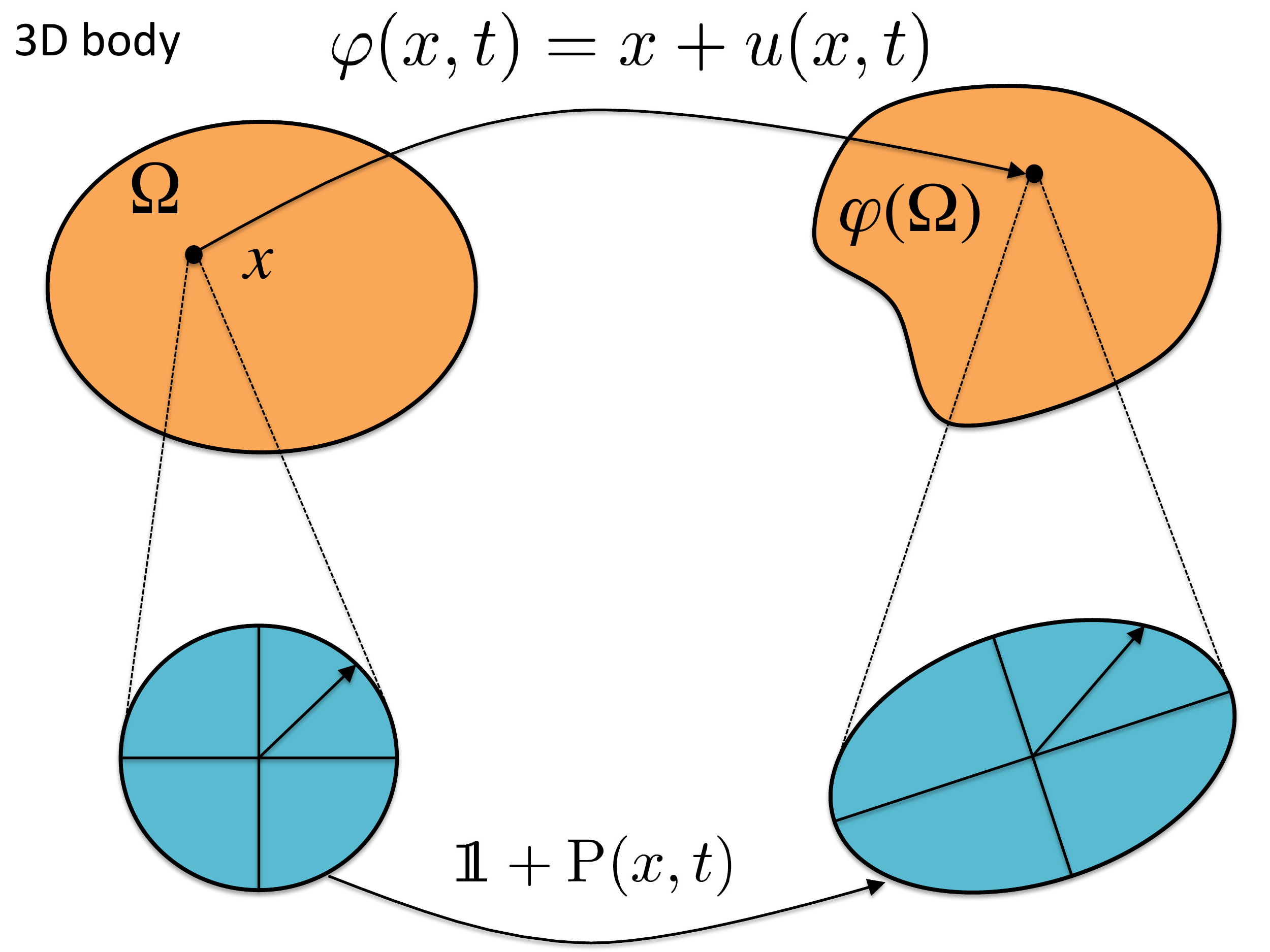}}{\caption{Enriched kinematics for a micromorphic continuum. The macroscopic deformation of the body $\Omega\subset\R^{3}$ is described by $\varphi:\,\Omega\subset\R^{3}\rightarrow\R^{3}$. In each macroscopic material point $x\in\Omega$ there is a substructure attached. This substructure has the possibility to shear, stretch and shrink and is described by an affine mapping $\mathds{1}+\p$. Decisive is {the constitutive choice of the strain energy density which couples the} macroscopic displacement $u$ and the micro-distortion $\p$. Our new relaxed micromorphic model introduces the weakest possible {constitutive} coupling still giving a well-posed model.\newline \newline   \label{fig:kinematics}}}
 \end{figure}

 In the micromorphic model, it is the kinematics which is enriched by introducing an additional field of \textbf{non-symmetric micro-distortions} $\p:\Omega\subset\R^{3}\rightarrow\R^{3\times3}$, beyond the classical macroscopic displacement $u:\Omega\subset\R^{3}\rightarrow\R^{3}$ {(see Fig. \ref{fig:kinematics})}. Then, a \textbf{non-symmetric elastic (relative) distortion} $e=\nablau-\p$ can be defined and the modeling proceeds by obtaining the constitutive relations linking elastic-distortions to stresses and by postulating a balance equation for the micro-distortion field $\p$. All such steps might be preferably done in a variational framework, involving the third order curvature tensor (the micro-distortion gradient) $\nabla \hspace{-0.1cm}\p$, so that only energy contributions need to be defined a priori. For the dynamic case, one adds in the Hamiltonian the so-called micro-inertia density contributions, acting on the time derivatives of micro-distortion terms $\p_{,t}$.  
 
In principle, the modeling framework for the micromorphic approach had been completed by Eringen, Mindlin, in the references already cited, and Germain \cite{germain1973method}. {Mindlin and Eringen also provided} extensions {of the micromorphic model} to anisotropy {even if such anisotropic models are almost impossible to be applied to real cases, due to the impressive number of coefficients provided (498 coefficients in the general anisotropic case)}.

The existence and uniqueness questions for the linear micromorphic model have been completely settled both for the static and dynamic case, based on the assumption of \textbf{uniform positive definiteness} of the appearing constitutive elasticity tensors. However, the over-reliance on uniform positive definiteness, we believe, has blinded the eyes for the real possibilities inherent in the micromorphic model. These possibilities have been consistently overlooked until very recently, when, in a series of articles \cite{neff2014unifying,madeo2015wave,ghiba2014relaxed,madeo2014band,neff2015relaxed}, we have introduced the novel concept of \textbf{relaxed micromorphic continuum}. {This model provides a drastic reduction of the number of constitutive coefficients with respect to Mindlin-Eringens's one while remaining well-posed. }
\medskip

{Unlike Mindlin-Eringen's model,} the relaxed model mainly works with \textbf{symmetric elastic (relative) strains} $\varepsilon_{e}:=\sym \left(\nablau-\p\right)$, so that standard 4$^{th}$ order symmetric elasticity tensors can be used in order to define elastic stresses. {Moreover,} regarding the curvature, the relaxed model considers the \textbf{second order dislocation-density tensor} $\alpha=-\Curl \p$ instead of the third order curvature tensor $\nabla\hspace{-0.1cm} \p$ with the effect (among others) that the description of the anisotropy of curvature only needs 4$^{th}$ order tensors, instead of 6$^{th}$ order ones.

A fundamental contribution of the relaxed micromorphic model is given by the fact that well-posedness results have been proven \cite{neff2014unifying} also for the case where the strain energy density violates strict positive-definiteness{\footnote{It has to be noted that our new approach is only formally included in the standard Mindlin-Eringen micromorphic model since we consistently give up uniform positive-definiteness in the elastic distortion $e$ and the curvature tensor $\nabla\p$ which are instead strictly requested in the standard model in order to have well-posedness. For example, controlling only the elastic strain $\varepsilon_{e}=\sym \left(\nablau-\p\right)$ in the energy does not locally control the elastic distortion $e=\nablau-\p$ and working with $\Curl \p$ does not control the curvature $\nabla\hspace{-0.1cm} \p$.}}. In other words, even if the relaxed micromorphic model can be apparently seen as a particular case of the Mindlin-Eringen model by suitably setting  some constitutive parameters of their model to zero (see \cite[p. 555]{madeo2015wave}), such choice is not acceptable in the Mindlin-Eringen setting due to the loss of positive-definiteness of the energy. Nevertheless, it is exactly this feature which makes the relaxed micromorphic model unique for the description of a wealth of unorthodox material behaviors. The existence results proposed in \cite{neff2014unifying}, { as well as the drastic reduction of the number of the introduced elastic coefficients, allowed us to open the way to the application of the relaxed micromorphic model to cases of real interest.

Indeed, the relaxed micromorphic model} has already been a source of inspiration for researchers working on granular materials \cite{misra2015granular}.\footnote{ Although delighted by the fact of understanding that the relaxed micromorphic model might be of use for granular mechanics, we believe that some complements of information must be given in order to interpret the results of \cite{misra2015granular} in the clearest possible way. In \cite{misra2015granular} the authors use micro-macro upscaling techniques for granular assemblies  arriving to a standard Mindlin-Eringen type model at the homogenized scale (see \cite[p.224, eq. (43)]{misra2015granular}. The authors observe that: ``\textsl{remarkably, the nonzero components in Mindlin's stiffness tensors are the same as the non-zero components derived from the present model}''. Then, the authors present in equation (66) a constitutive choice of the microscopic parameters which goes in the sense of setting to zero the parameters of Mindlin's model in order to get close to the relaxed micromorphic model. Such constitutive choice is not justified neither by telling that the scope is to recover the relaxed micromorphic model nor on clear microscopic-based arguments that would shed additional light on the understanding of microstructure-related effects.

Afterwards, the authors present \cite[p.231, Fig. 5]{misra2015granular} two parametric studies on the parameters $\beta_{mM}$ and $\beta_{sM}$. The parameter $\beta_{mM}$ is an analogous of the Cosserat couple modulus $\mc$ and is once again seen to be determinant for the onset of band gaps. On the other hand, the parameter $\beta_{sM}$ is the one that, being non vanishing, still makes a difference between Mindlin-Eringen's and our relaxed model. The authors then present a parametric study letting  $\beta_{sM}$ to zero, which indeed means that they are recovering the relaxed micromorphic model as a limit case. Nevertheless, except some mostly confusing sentences referring to our paper \cite{madeo2015wave}, such fundamental observation are not made at any point of the paper \cite{misra2015granular}.

It should be clearly stated that, by means of the proposed parametric study, they are trying to approach the relaxed micromorphic model and that, although the corresponding choice of the parameters is not allowed in Mindlin-Eringen theory, the well-posedness is still guaranteed. Moreover, it should have been clearly stated that the \textbf{relaxed micromorphic model} is the \textbf{only generalized {non-local} continuum model}, among those currently used, which is able to predict \textbf{complete frequency band-gaps}\cite{ghiba2014relaxed,madeo2014band,madeo2015wave,neff2015relaxed}.

Finally, and this would be for us the main advancement related to the paper \cite{misra2015granular}, a clear microscopic-based interpretation of the fact of setting to zero the opportune parameters in Mindlin's theory would be necessary in further works since it is not currently done in \cite{misra2015granular}. Of course, the fact of setting to zero some macroscopic parameters leads to some conditions on some microscopic parameters, but which is the physical interpretation of such conditions on micro-parameters?

In summary, the same goal of clarity that we try to pursue in this paper should be, from our point of view, shared by the highest possible number of researchers in order to proceed in the direction of a global advancement of knowledge.} { Moreover, the clear and transparent application of the relaxed micromorphic model in the isotropic case has recently been successfully achieved for the description of band-gap metamaterials (see \cite{madeo2016first,madeo2016modeling}).
	
	As a matter of fact, the isotropic relaxed micromorphic model has proven its ability to fit the dispersion curves of phononic crystals for large windows of frequencies and wavelengths, arriving down to wavelengths which are comparable to the size of the unit cell. The most interesting aspect of the description of such metamaterials via the relaxed micromorphic model is undoubtedly that of predicting their macroscopic dynamical response through the introduction of \textbf{few macroscopic elastic coefficients} which are \textbf{independent of the frequency}.
	
	This means that the coefficients of the relaxed micromorphic model can be seen as \textbf{true material parameters}, exactly as it is the case for the Young modulus and the Poisson ratio when dealing with classical materials.
	
	Of course, in order to extend the range of applicability of the relaxed model to a wider class of actual metamaterials, the model must be generalized to the anisotropic setting. This generalization is the principal aim of the present work. 
}
\medskip

In this paper, we want to present such an approach to anisotropy for the relaxed micromorphic model. Our modeling perspective is to simplify as much as possible, and indeed to reduce to an essential minimum, the bewildering possibilities of the standard micromorphic model. Indeed, there is no point in exclaiming happily that the standard micromorphic model has more than 1000 constitutive coefficients which need to be determined. The true aim of modeling should consist of the opposite: discard all unclear complications without compromising the essence of the model. We believe that the relaxed micromorphic model is just going in this direction, thereby opening the way {to transparent} experimental campaigns for the determination of the remaining fewer extra parameters.

The plan of the paper is as follows.
\begin{itemize}
	\item  We first recall the standard micromorphic model and contrast it with our new relaxed model. We also show that our relaxed micromorphic model supports a clear group-invariant framework, opening the way to {the introduction of} anisotropy classes.
	\item We present our favored description of anisotropy regarding the higher order contribution in $\Curl \p$. Thereby we split $\Curl \p$ as $\sym \Curl \p+\skew \Curl \p$ and let a classical fourth order tensor act only on $\sym \Curl \p\in\Sym(3)$ together with another tensor with only 6 parameters acting on $\skew \Curl \p\in\so(3)$.
	\item We consider the long-wavelength limit (characteristic length $L_{c}\rightarrow 0$) which must coincide with a linear elastic model that has lost any characteristic length (sometimes  called internal variable model). From this hypothesis, we are able to relate coefficients of the micromorphic scale to the macroscopic ones. The result is a convincing homogenization formula for all considered anisotropy classes. This is also done using classical Voigt-notation in order to facilitate future applications. {As already stated, this homogenization formula relating micro and macro parameters is one of the main results of the present work, since it opens the way to the application of the model to actual metamaterials via the realization of standard mechanical tests on ``large'' specimens.}
	\item We study the format of a possible anisotropic local rotational coupling term acting on $\skew \left(\nablau-\p\right)$. In this respect we also investigate some possibilities of approximating an anisotropic coupling by an isotropic one.
	\item We consider the formal limit $L_{c}\rightarrow \infty$ and show that it corresponds to a ``zoom'' into the micro-structure. Our relaxed model supports also a clear interpretation for that regime. 
	\item We end our paper by showing that the standard Mindlin-Eringen micromorphic model does not support the clear relation between macroscopic and microscopic elasticity moduli which is instead provided by our simplified anisotropic relaxed model.
\end{itemize}

\section{Notational agreement}

Throughout this paper Latin subscripts take the values $1,2,3$ while Greek subscripts take the values $1,2,3,4,5,6$ and we adopt the Einstein convention of sum over repeated indices if not differently specified.

We denote by $\R^{3\times3}$ the set of real $3\times3$ second order tensors and  by $\R^{3\times3\times3}$ the set of real $3\times3\times3$ third order tensors. The standard Euclidean scalar product on $\R^{3\times3}$ is given by $
\langlenew{X},{Y}\ranglenew_{\R^{3\times3}}=\tr({X \x Y^{T}}) $
and, thus, the Frobenius tensor norm is $\|{X}\|^{2}=\langlenew{X},{X}\ranglenew_{\R^{3\times3}}$. Moreover, the identity tensor on $\R^{3\times3}$ will be denoted by $\mathds{1}$, so that $\tr({X})=\langlenew{X},{\mathds{1}}\ranglenew$. We adopt the usual abbreviations of Lie-algebra theory, i.e.:
\begin{itemize}
	\item $\Sym(3):=\{X\in\R^{3\times3}\;|X^{T}=X\}$ denotes the vector-space of all symmetric $3\times3$ matrices
	\item $\so(3):=\{X\in\R^{3\times3}\;|X^{T}=-X\}$
	is the Lie-algebra of skew symmetric tensors 
	\item $\mathfrak{sl}(3):=\{X\in\R^{3\times3}\;|\tr({X})=0\}$
	is the Lie-algebra of traceless tensors
	\item $\R^{3\times3}\simeq\mathfrak{gl}(3)=\{\mathfrak{sl}(3)\cap\Sym(3)\}\oplus\so(3)\oplus\R\!\cdot\!\mathds{1}$ is the \emph{orthogonal Cartan-decomposition of the Lie-algebra}
\end{itemize} 
For all $X\in\R^{3\times3}$, we consider the decomposition 
\begin{align}X=\dev\sym X+\skew X+\frac{1}{3}\mathrm{tr}(X)\,\mathds{1}\end{align}
 where:
\begin{itemize}
	\item $\sym\, X=\frac{1}{2}(X^{T}+X)\in\Sym(3)$ is the symmetric part,
	\item $\skew\, X=\frac{1}{2}(X-X^{T})\in\so(3)$ is the skew-symmetric part,
	\item $\dev\, X=X-\frac{1}{3}\tr(X)\,\mathds{1}\in\mathfrak{sl}(3)$ is the deviatoric part .
\end{itemize}

Throughout all the paper, we denote:
\begin{itemize}
	\item the sixth order tensors  $\Ls:\R^{3\times3\times3}\rightarrow\R^{3\times3\times3}$ by a hat
	\item the fourth order tensors  $\Cfourth:\R^{3\times3}\rightarrow\R^{3\times3}$ by overline
	\item without superscripts, i.e.$\,\Csym$, the classical fourth order tensors acting only on symmetric matrices \\ $\Csym:\Sym(3)\rightarrow\Sym(3)$ or skew-symmetric ones $\Cc:\so(3)\rightarrow\so(3)$
	\item the second order tensors  $\widetilde{\Csym}:\R^{6}\rightarrow\R^{6}$ or $\widetilde{\Csym}:\R^{3}\rightarrow\R^{3}$ appearing as elastic stiffness by a tilde.
\end{itemize}

We denote by $\Cfourth\X X$ the linear application of a $4^{th}$ order tensor  to a $2^{nd}$ order tensor and also for the linear application of a  $6^{th}$ order tensor $\Ls$ to a $3^{rd}$ order tensor. In symbols:
\begin{align}
\left(\Cfourth \X X\right)_{ij}=\Cfourth_{ijhk} X_{hk}\,,\qquad \left(\Ls \X A\right)_{ijh}=\Ls_{ijhpqr} A_{pqr}\, .
\end{align} 
The operation of simple contraction between tensors of suitable order is denoted by a central dot, for example:
\begin{align}
\left(\Csecond \x v\right)_{i}=\Csecond_{ij} v_{j}\,, \qquad \left(\Csecond \x X\right)_{ij}=\Csecond_{ih} X_{hj}\, .
\end{align}

Typical conventions for differential operations are implied, such as a comma followed by a subscript to denote the partial derivative with respect to the corresponding Cartesian coordinate, i. e. $\left(\cdot\right)_{,j}=\frac{\partial(\cdot)}{\partial x_j}$.

Given a skew-symmetric matrix $\overline{A}\in \so(3)$ we consider:
\begin{align}
\overline{A}=\left( \begin{matrix}
0 & \overline{A}_{12} & \overline{A}_{13} \\-\overline{A}_{12} & 0 & \overline{A}_{23} \\ -\overline{A}_{13} & -\overline{A}_{23} & 0 
\end{matrix}\right),\qquad\qquad 
\axl\left(\overline{A}\right)=(-\overline{A}_{23},\overline{A}_{13},-\overline{A}_{12})^T.
\end{align}
ore equivalently in index notation:
\begin{align}
\left[\axl\left(\overline{A}\right)\right]_{k}=-\frac{1}{2} \epsilon_{ijk}\, \overline{A}_{ij}=\frac{1}{2} \epsilon_{kij}\, \overline{A}_{ji}\,, \label{axl}
\end{align}
where $\epsilon$ is the Levi-Civita third order permutation
tensor.

\section{A review on the micromorphic approach}
{
In this section we recall the general anisotropic setting of classical Mindlin-Eringen micromorphic elasticity, as well as that of relaxed micromorphic elasticity. We show that, given its intrinsic formulation, the relaxed micromorphic model features 93 coefficients instead of Mindlin/Eringen 498.

In subsection \ref{symmetry}, a further reduction of coefficient is proposed for those cases in which one wants to feature a symmetric stress. 

In subsection \ref{curvature}, it is shown that the most general form of the relaxed curvature energy (in the anisotropic setting) which satisfies  certain additional invariance requirements features 21+7=28 coefficients instead of the 378 featured by the classical Mindlin-Eringen model. Moreover some further simplification of the curvature energy are proposed, up to arriving to the isotropic case in which the curvature energy only shows 3 coefficients. As a matter of fact, we propose a consistent framework for the definition of the curvature energy of the relaxed micromorphic model which is fully consistent with invariance arguments. Such clear theoretical framework is of primordial importance for the introduction of suitable constitutive expressions for the curvature energy. Nevertheless, it is likely that, in a first instance, non-local effects in real metamaterials can be controlled via the introduction of very few characteristic lengths. For this reason, the maximum generality of the anisotropic setting for the curvature could find effective applications only in a second instance, when the most important step of the identification of the elastic coefficients $\Ce$, $\Ch$ and $\Cc$ will be achieved on a suitable class of targeted metamaterials.

In subsection \ref{microinertia}, the general anisotropic setting for the kinetic energy to be used in the relaxed micromorphic model is provided. This step is strongly complementary to the constitutive choice for the static case featured by equation \eqref{eq:Ener3}. Indeed, if some deformation mechanism are introduced in the definition of the strain energy densities, analogous inertiae must be introduced in the kinetic energy to have a well-posed problem in the dynamical case. This step is essential to securely proceed towards controllable applications on actual metamaterials subjected to dynamical loading.
}
\subsection{The standard Mindlin-Eringen model}
The elastic energy of the general anisotropic centro-symmetric micromorphic model in
the sense of Mindlin-Eringen (see \cite{mindlin1964micro} and \cite[p. 270, eq. 7.1.4]{eringen1999microcontinuum}) can be represented
as:
\begin{align}
W= & \underbrace{\frac{1}{2}\langlenew\Coe \X \left(\nablau-\p\right),\left(\nablau-\p\right)\ranglenew_{\R^{3\times 3}}}_{\mathrm{{\textstyle full\ anisotropic\ elastic-energy}}}+\underbrace{\frac{1}{2}\langlenew\Ch \X\sym \p,\sym \p\ranglenew_{\R^{3\times 3}}}_{\mathrm{\textstyle micro-self-energy}}\label{eq:EnerEringen}\\
 &+\underbrace{\frac{1}{2} \langlenew \E \X \left(\nablau- \p\right), \sym \p\ranglenew_{\R^{3\times 3}}}_{\mathrm{\textstyle anisotropic\ cross-coupling}}+\underbrace{\frac{\mLc}{2} \langlenew \Ls_{\text{aniso}} \X \nabla\hspace{-0.1cm}  \p,\nabla\hspace{-0.1cm} \p\ranglenew_{\R^{3\times 3\times 3}}}_{\mathrm{\textstyle full\ anisotropic\ curvature}}\,,\nonumber 
\end{align}
where $\Coe:\R^{3\times3}\rightarrow\R^{3\times3}$
is a $4^{th}$ order micromorphic elasticity tensor which has at most 45 independent coefficients and which acts on the \textbf{non-symmetric elastic distortion} $e=\nablau-\p$ and $\E:\R^{3\times3}\rightarrow\Sym(3)$ is a $4^{th}$ order cross-coupling tensor with the symmetry $\left(\E\right)_{ijkl}=\left(\E\right)_{jikl}$ having at most 54 independent coefficients. The fourth order tensor $\Ch:\Sym(3)\rightarrow\Sym(3)$
has the classical 21 independent coefficients of classical elasticity, while $\Ls_{	\text{aniso}}:\R^{3\times 3\times 3}\rightarrow\R^{3\times 3\times 3}$ is a $6^{th}$ order tensor that shows an astonishing 378 parameters. The parameter $\mu>0$ is a typical shear modulus and $L_{c}>0$ is one characteristic length, while $\Ls_{\text{aniso}}$ is, accordingly, dimensionless. Here, for simplicity, we have assumed just a decoupled format of the energy: mixed terms of strain and curvature have been discarded by assuming \textbf{centro-symmetry}. Counting the number of coefficients we have $45+21+54+378=498$ independent coefficients.

If we assume an isotropic behavior of the curvature we obtain:
\begin{align}
W= & \underbrace{\frac{1}{2}\langlenew\Coe \X \left(\nablau-\p\right),\left(\nablau-\p\right)\ranglenew_{\R^{3\times 3}}}_{\mathrm{{\textstyle full\ anisotropic\ elastic-energy}}}+\underbrace{\frac{1}{2}\langlenew\Ch \X\sym \p,\sym \p\ranglenew_{\R^{3\times 3}}}_{\mathrm{{\textstyle micro-self-energy}}}\label{eq:EnerEringenIso}\\
&+\underbrace{\frac{1}{2} \langlenew \E \X \left(\nablau- \p\right),\sym \p\ranglenew_{\R^{3\times 3}}}_{\mathrm{\textstyle anisotropic\ cross-coupling}}+\underbrace{\frac{\mLc}{2} \langlenew \Ls_{\text{iso}} \X \nabla\hspace{-0.1cm} \p,\nabla\hspace{-0.1cm} \p\ranglenew_{\R^{3\times 3\times 3}}}_{\mathrm{\textstyle{isotropic\ curvature}}}\,,\nonumber 
\end{align}
where the $6^{th}$ order tensor $ \Ls_{\text{iso}}$ has still 11 independent non-dimensional constants \cite{eringen1999microcontinuum}. This can be explained considering that the general isotropic $6^{th}$ order tensor has 15 coefficients which, considering that in a quadratic form representation we can assume a major symmetry of the type $\Ls_{ijklmn}=\Ls_{lmnijk}$, reduce to 11  (see \cite{spencer1971theory,monchiet2011inversion}).\footnote{{The 11 coefficients of the curvature in the isotropic case reduce to 5 in the particular case of second gradient elasticity (see \cite{dellisola2009generalized}) which is obtained from a micromorphic model by setting $\p=\nablau$}.} On the other hand, the local energy has 7 independent coefficients in the isotropic case: $\Coe$ has 3, $\Ch\sim2$, $\E\sim 2$ adding up to the usual 18 constitutive coefficients to be determined in the isotropic case. 

One of the major obstacles in using the micromorphic approach for specific materials is the impossibility to determine such multitude of new material coefficients. Not only is the huge number a technical problem, but also the interpretation of coefficients is problematic \cite{chen2003determining,chen2003connecting,chen2004atomistic}. Some of these coefficients are size-dependent while others are not. A purely formal approach, as is often done, cannot be the final answer.

\subsection{The relaxed micromorphic model}
Our novel relaxed micromorphic model endows Mindlin-Eringen's representation
with more geometric structure. Since $\E$ is difficult to interpret, it is discarded right-away. Nevertheless, the structure of the model continues to be very rich. We write:
\begin{align}
W= & \underbrace{\frac{1}{2}\langlenew\Ce \X \sym\left(\nablau-\p\right),\sym\left(\nablau-\p\right)\ranglenew_{\R^{3\times3}}}_{\mathrm{{\textstyle anisotropic\ elastic-energy}}}+\frac{1}{2}\underbrace{\langlenew\Ch \X\sym \p,\sym \p\ranglenew_{\R^{3\times3}}}_{\mathrm{\textstyle micro-self-energy}}\label{eq:Ener3}\\
 & \ 
+\underbrace{\frac{1}{2}\langlenew\Cc \X \skew\left(\nablau-\p\right),\skew\left(\nablau-\p\right)\ranglenew_{\R^{3\times3}}}_{\begin{array}{c}
\mathrm{\textstyle invariant\ local\ anisotropic} \\
\mathrm{\textstyle rotational\ elastic\ coupling\ }	
	\end{array}
} 
+\underbrace{\frac{\mLc}{2} \langlenew \Lf_{\text{aniso}} \X \, \Curl \p,\Curl \p\ranglenew_{\R^{3\times3}}}_{\mathrm{\textstyle curvature}}\,.
\nonumber 
\end{align}
The second order tensor $\alpha:=-\Curl \p$ is usually called the \textbf{dislocation density tensor}.\footnote{The dislocation tensor is defined as $\alpha_{ij}=-\left(\Curl \p \right)_{ij}=-\p_{ih,k}\epsilon_{jhk}$, where $\epsilon$ is the Levi-Civita tensor.} Here  $\Ce,\ \Ch~:~\Sym(3)~\rightarrow~\Sym(3)$
are both classical $4^{th}$ order elasticity tensors \textbf{acting on symmetric second order tensors} only: $\Ce$ acts on the \textbf{symmetric elastic strain} $\varepsilon_{e}:=\sym \left(\nablau-\p\right)$ and $\Ch$ acts on the \textbf{symmetric micro-strain} $\sym \p$ and both map to symmetric tensors. The tensor  $\Cc:\so(3)\rightarrow\so(3)$ is a $4^{th}$ order tensor that acts only on skew-symmetric matrices and yields only skew-symmetric
tensors and $\Lf_{\text{aniso}}:\R^{3\times 3}\rightarrow\R^{3 \times 3}$ is a dimensionless $4^{th}$ order tensor with at most 45 constants.  
Counting coefficients we now have 21+21+6+45=93, instead of Mindlin-Eringen's
498 coefficients. The main advantage at this stage is that our $\Ce$, unlike $\Coe$, possesses all the symmetries that are peculiar of the classical elasticity tensors acting on $\sym \nablau$.

The large number of isotropic constants in the standard Mindlin-Eringen model has always been of concern. Previous attempts to endow the Mindlin-Eringen model with more structure include Koh's \cite{koh1970special,parameshwaran1973wave} so-called micro-isotropy postulate which requires, among others, that $\sym\, \sigma$ is an isotropic function of $\sym \nablau$ only. This reduces the number of isotropic coefficient also to 5 (similarly to our relaxed model) but {the fact of} connecting $\sym\, \sigma$ to $\sym \nablau$ only {cannot be considered to be a well-grounded hypothesis}.

Considering the energy in equation \eqref{eq:Ener3}, the resulting elastic stress is:
\begin{align}
\sigma\left(\nablau,\p\right)=\,\Ce \X \sym\left(\nablau-\p\right)+\Cc \X \skew\left(\nablau-\p\right) & ,\label{eq:SigmaMicro}
\end{align}
which is solely related to elastic distortions $e=\nablau-\p$. {One of the main results of the present paper is to provide a simple but effective homogenization formula which relates the elastic tensors $\Ce$ and $\Ch$ to the macroscopic elastic properties of the considered medium that will be encoded in the effective elastic tensor $\C$.}

{The derivation of the macroscopic consistency condition we propose in the present paper is of primary importance for an effective application of the proposed model to cases of real interest.}

{Indeed, the basic idea is that of considering a sample of a specific microstructured material which is large enough to let the effect of the underlying microstructure being negligible. On this large sample, standard mechanical tests can be performed to allow for the unique determination of the elastic coefficients $\C$.}

{The existence of our formula relating $\C$ (which is well known) to $\Ce$ and $\Ch$ (which are still unknown), allows to further reduce the number of coefficients that need to be determined to unequivocally characterize the mechanical behavior of microstructured materials.}

This unique feature of our relaxed model gives again more credibility to the relaxed approach {by opening the way to a} clear experimental campaign to determine some of the new micromorphic elastic constants. 

In the general anisotropic micromorphic model initially proposed by
Mindlin-Eringen \cite{eringen1964nonlinear} the question of parameter identification has already been treated. However, the resulting interpretation of the material constants, as well as their connection to the classical anisotropy formulation of linear elasticity, is still not settled satisfactorily{, and presumably impossible.}

As already seen, in our relaxed model the complexity of the general micromorphic model has been decisively reduced, featuring basically only symmetric strain-like variables and the $\Curl$ of the micro-distortion $\p$. However, the relaxed model is still general enough to include the full micro-stretch as well as the full Cosserat micro-polar model, see \cite{neff2014unifying}. Furthermore, well-posedness results for the static and dynamic cases have been provided in \cite{neff2014unifying} making decisive use of recently established coercive inequalities, generalizing Korn's inequality to incompatible tensor fields  \cite{neff2015poincare,neff2002korn,neff2012maxwell,neff2011canonical,bauer2014new}.

{Furthermore, certain limiting cases of the anisotropic relaxed micromorphic model give as a result other micromorphic models, e.g. the Cosserat model, the micro-dilation theory, the micro-incompressible micromorphic model, the micro-stretch theory and the microstrain model) as it is shown in Appendix \ref{limiting}. Instead, the second gradient model cannot be found as a limiting case differently from what happens in the Eringen Mindlin micromorphic model, see Appendix \ref{1D} for the one dimensional case.}

 \subsubsection{ Possible symmetry of the relaxed micromorphic stress \label{symmetry}}
 
In this subsection, we {recall some arguments that allow the possibility of featuring a symmetric stress tensor for the relaxed micromorphic model by setting the 6 components of the tensor $\Cc$ to be vanishing.} Considering the scalar product $\langlenew X,Y\ranglenew=\tr(X \x Y^{T})$, we start by noticing that{, given the definition of the fourth order tensors $\Ce$ and $\Cc$, they respect a generalized version of the \textbf{orthogonal decomposition} of second order tensors ($X=\sym X\oplus\skew X$),} in the sense that:
 \begin{align}
 \sym\,[\Ce \X \sym X+\Cc \X \skew X]&=\Ce \X \sym X, \label{eq:Decom}\\
 \skew\,[\Ce \X \sym X+\Cc \X \skew X]&=\Cc \X \skew X\,.\nonumber
 \end{align}
 We recall that the elastic stress {of the relaxed micromorphic model} is:
 \begin{align}
 \sigma\left(\nablau,\p\right)=\,\Ce \X \sym\left(\nablau-\p\right)+\Cc \X \skew\left(\nablau-\p\right) & ,
 \end{align}
{so that} skew-symmetry of the elastic stress $\sigma$ is entirely controlled by the rotational coupling tensor $\Cc$ since{, relying on formulas \eqref{eq:Decom},} we have
\begin{align}\skew \,\sigma=\skew\left[\Ce \X \sym\left(\nablau-\p\right)+\Cc \X \skew\left(\nablau-\p\right)\right]=\Cc \X \skew\left(\nablau-\p\right) & .\label{eq:SkewSigmaMicro}\end{align}
For a positive definite coupling tensor $\Cc$, we note that skew-symmetric stresses$\skew\,\sigma\neq0$ occur if and only if $\skew\left(\nablau-\p\right)\neq0$. 
\begin{framed} 
	If $\Cc\equiv0$, the elastic Cauchy stress $\sigma$ satisfies \textbf{Boltzmann's axiom of symmetry of force stresses}. In addition, for $\Cc\equiv0$, the \textbf{elastic distortion} $e=\nablau-\p$ can be non-symmetric, while the elastic stress $\sigma$ remains symmetric.\footnotemark  \end{framed}\footnotetext{\label{footKroner}Therefore, using $\Cc=0$ is similar to the Reuss-bound approach in homogenization theory in which the guiding assumption is that the stress fields are taken constant but fluctuations in strain are allowed. Here, analogously, we would assume symmetric stresses $\sigma$ but non-symmetric distortion-fluctuations in $e=\nablau-\p$. Voigt (see  \cite[p.596]{voigt1908lehrbuch}) already discussed non-symmetric states of distortion. However, we can supply some further support for using $\Cc\equiv 0$. Indeed as  Kr\"oner notes  \cite{kroner1955fundamentale}, \textsl{``asymmetric stress tensors only come under consideration when a distribution of rotational moments acts upon the body externally, which is excluded here. The question of whether the (...) rotations produces stresses can also be answered. We must first exclude asymmetric stress tensors, since they contradict the laws of equilibrium in the theory of elasticity''. Furthermore, Kunin \cite[p. 21]{kunin2012elastic} states the following theorem: \textsl{in the nonlocal theory of a linear elastic medium of simple structure with finite action-at-a-distance, it is always possible to introduce a symmetric stress tensor and an energy density, which can be expressed in terms of stress and strain in the usual way}.}}
 In \cite{romano2016micromorphic} the authors have introduced the original and important notion of \textbf{non-redundant strain measures} in the micromorphic continuum.  As it turns out, the relaxed micromorphic model with zero rotational coupling tensor $\Cc\equiv0$ is \textsl{a non-redundant micromorphic formulation}. Conversely, the standard Mindlin-Eringen model remains \textbf{redundant}, as does the linear Cosserat model.
 
With Boltzmann's axiom, which is in sharp contrast to standard micromorphic models, the model would feature symmetric force-stress tensors. Such an assumption has been made, for example, by Teisseyre \cite{teisseyre1973earthquake,teisseyre1974symmetric} in his model for the description of seismic wave propagation phenomena (for the use of micromorphic models for earthquake modeling see also the discussion in \cite{nagahama2000micromorphic}).

It must also be observed that the relaxed micromorphic model can be used with $\Cc$ positive semi-definite or indeed zero (in the isotropic case $\mc=0$), while we {always} assume that $\Ce$, $\Ch$ (and later $\C$) are strictly positive definite tensors. Assuming that $\Ce$ and $\Ch$ are positive definite tensors means that:
\begin{align}
\exists \, c_{e}^{+} >0: \, \forall S \in \text{Sym}(3):\quad \langlenew\Ce \X S,S\ranglenew_{\R^{3\times3}} \geq c_{e}^{+} \|S \|^2_{\R^{3\times3}}.
\end{align}
 In sharp contrast to the standard Mindlin-Eringen format, we assume for the rotational coupling tensor $\Cc$ only positive semi-definiteness, i.e:
 \begin{align}
 \forall \overline{A} \in \so(3):\quad\langlenew\Cc \X \overline{A},\overline{A}\ranglenew_{\R^{3\times 3}} \geq 0.
 \end{align}
 As already noted, this allows the rotational coupling tensor $\Cc$ to vanish, in which case the relaxed micromorphic model is \textbf{non-redundant} \cite{romano2016micromorphic}.\footnote{In Misra et al. \cite{misra2015granular} the rotational coupling $\Cc$ is related to the tangential stiffness between grains. This is consistent with Shimbo's law \cite{shimbo1975geometrical} relating the rotational stiffness to the internal friction. We need to remark that friction is, strictly speaking, a dissipative effect outside purely elastic response.}
 
The reader might ask himself: how is it possible that the rotational coupling tensor $\Cc$ can be absent but the resulting model is still well-posed? This is possible because in that case, the skew-symmetric part of $\p$ is not controlled locally but as a result of the boundary value problem and boundary conditions.
 In this sense, allowing for $\Cc\equiv 0$ is one of the decisive new possibilities offered by the relaxed micromorphic model.
 
 However, in \cite{madeo2015wave} it has been shown that in the isotropic case ($\Cc=\mc \,\mathds{1}$) the presence of $\Cc$ allows to control the onset of band-gaps. In section \ref{sec:Mandel-Voigt-vector-notation} we discuss the possible forms that $\Cc$ may have for certain given anisotropy classes.

\subsubsection{Microscopic curvature \label{curvature}}

In the general micromorphic model, the curvature energy term is of the form:
\begin{align}
W_{\mathrm{curv}}=W_{\mathrm{curv}}(\nabla\hspace{-0.1cm} \p),
\end{align} 
In our relaxed framework, we assume that it depends only on the second order dislocation density tensor through:
\begin{align}
W_{\mathrm{curv}}^{\mathrm{relax}}=W_{\mathrm{curv}}^{\mathrm{relax}}(\Curl \p).\label{eq:encur}
\end{align} 
First, we remark here that the relaxed micromorphic curvature expression {can also be written as}:
\begin{align}
\Curl \p=-\Curl \left(\nablau-\p\right)\,,
\end{align}
because $\Curl \p$ is invariant under $\p\rightarrow\p+\nabla \vartheta$, see \cite{neff2015relaxed}.

Now, we need to shortly discuss that such a reduced formulation is fully able to be treated in an invariant setting. 
To this end, let $\p\,\mathrm{:}\,\Omega \subset \R^{3}\rightarrow \R^{3\times3}$ be the micro-distortion field {to which we apply} the following coordinate transformation (generating the so-called Rayleigh-action on it \cite{auffray2013algebraic}):
\begin{align}
\p^{\#}(\xi):=Q \x \left[\p (Q^{T}\x \xi)  \right] \x Q^T\,,\qquad x=Q^{T} \x \xi,\label{eq:rot}
\end{align}
for given $Q\in \mathrm{SO}(3)$. Transforming the displacement to a rotated reference and spatial configuration, we have:
\begin{align}u^{\#}(\xi):=Q \x u\,(Q^T \x \xi)\,,\qquad\nabla_{\xi}u^{\#}(\xi)=Q\x\nabla_{x}u(Q^{T} \x \xi) \x Q^T\,,
\end{align}
thus, we require that $\p$ transforms as $\nablau$ under simultaneous rotations of the reference and spatial configurations. Then it can be shown\cite{munch2016rotational} that:
\begin{align}
\Curl_{\xi}\p^{\#}(\xi)=Q \x[\Curl_{x}\p  (Q^{T}\x\xi)]\x Q^{T}.\label{eq:rot2}
\end{align}
For the description of anisotropy in the curvature energy we require \textbf{form-invariance} of expression \eqref{eq:encur} under the transformation \eqref{eq:rot} with respect to all rotations $Q\in\mathcal{G}$-material symmetry group. Taking \eqref{eq:rot2} into account, this means
\begin{align}
\forall Q\in\mathcal{G}-\text{material\ symmetry\ group}:\qquad W_{\mathrm{curv}}^{\mathrm{relax}}(Q^{T}\x\Curl \p\x Q)=W_{\mathrm{curv}}^{\mathrm{relax}}(\Curl \p).\label{Invariance}
\end{align} 
In the same spirit as done with the local energy terms, a first simplification of the curvature expression{, which is consistent with the invariance condition \eqref{Invariance}} is given by:
\begin{align}
W_{\mathrm{curv}}^{\mathrm{relax}}\left(\Curl \p\right)=&\,\frac{\mLc}{2}\Big[\langlenew\Lsym_{e} \X \sym \Curl \p,\sym \Curl \p \ranglenew + \langlenew \Lsym_{c} \X \skew \Curl \p,\skew \Curl \p \ranglenew  \Big]\,.
\end{align}
Here,  $\Lsym_{e}:\Sym(3)\rightarrow\Sym(3)$ is a classical, positive definite elasticity tensor with at most 21 independent (non-dimensional) coefficients  and  $\Lsym_{c}:\so(3)\rightarrow\so(3)$  is a positive definite tensor with at most 6 independent (non-dimensional) coefficients. Taking isotropy into account, the total number of coefficients reduces to 3, while in the cubic case we have 4 coefficients.

We can think of another reduction of the curvature expression which is fully consistent with group-invariance requirements. Let:
\begin{align}
W_{\mathrm{curv}}^{\mathrm{relax}}=\widehat{W}_{\mathrm{curv}}^{\mathrm{relax}}\left(\sym \Curl \p\right).\label{eq:encursym}
\end{align}
Considering the same transformation law \eqref{eq:rot} as before, the complete representation of anisotropy in terms of representation  \eqref{eq:encursym} is easy. Indeed, we may employ the classical format of the 4$^{th}$ order elasticity tensors to write:
\begin{align}
\widehat{W}_{\mathrm{curv}}^{\mathrm{relax}}\left(\sym \Curl \p\right)=\frac{\mLc}{2} \langlenew \Lsym_{e} \X \sym \Curl \p,\sym \Curl \p \ranglenew. \label{eq:encursym2}
\end{align}
Here $\Lsym_{e}:\Sym(3)\rightarrow\Sym(3)$ is a classical, positive definite elasticity tensor with at most 21 independent (non-dimensional) coefficients. The expression in \eqref{eq:encursym2} is certainly preferable for its simplicity for the treatment of {anisotropic curvatures}. However, it is not clear whether a formulation {based on} \eqref{eq:encursym2} can lead to mathematically well-posed results due to the current lack of a suitable coercive inequality for that case \cite{bauer2014new,bauer2016dev}. Our guess at the moment is that it should work for the micro-incompressible case, in which the constraint $\tr \p =0$ is appended. This case is reminiscent of gradient plasticity with plastic spin \cite{neff2011canonical,ebobisse2010existence,ebobisse2015existence} in which the micro-distortion $\p$ is identified with the plastic distortion.

As explained in detail in \cite{munch2016rotational}, isotropy of the curvature energy is tantamount to requiring \textbf{form-invariance} of expression \eqref{eq:encur} under the transformation \eqref{eq:rot}, i.e.:
\begin{align}
W_{\mathrm{curv}}^{\mathrm{relax}}\left(\Curl_{\xi} \p^{\#}(\xi)\right)=W_{\mathrm{curv}}^{\mathrm{relax}}\left(\Curl_{x} \p(x)\right).
\end{align}
Taking \eqref{eq:rot2} into account, isotropy of the curvature is satisfied if and only if:
\begin{align}
\forall Q \in \mathrm{SO}(3):\qquad W_{\mathrm{curv}}^{\mathrm{relax}}\left(Q\x \left(\Curl_{x} \p(x)\right) \x  Q^{T}\right)=W_{\mathrm{curv}}^{\mathrm{relax}}\left(\Curl_{x} \p(x)\right).
\end{align}
i.e. $W_{\mathrm{curv}}^{\mathrm{relax}}$ must be an isotropic scalar function.  We need to highlight the fact that $\Curl  \p$ is not just an arbitrary combination of first derivatives of $\p$ (and as such included in the standard Mindlin-Eringen most general anisotropic micromorphic format), but that the formulation in $\Curl \p$ supports a completely invariant setting, as seen in \cite{munch2016rotational}, \cite{neff2006cosserat}. Since $\Curl  \p$ is a second order tensor, it allows us to {discard the $6^{th}$ order tensors of classical Mindlin-Eringen micromorphic elasticity} and to work instead with $4^{th}$ order tensors whose anisotropy classification is much easier and well-known \cite{chadwick2001new}.

In general, if we consider an isotropic curvature term, we obtain the following representation:
\begin{align}
\frac{\mLc}{2} \langlenew \Lf_{\text{iso}} \X \, \Curl \p,\Curl \p\ranglenew_{\R^{3\times3}}=\frac{\mLc}{2} \left( \alpha_{1}\lVert \dev \sym \Curl  \p\rVert^2 + \alpha_{2}\lVert\skew \Curl  \p\rVert^2 + \alpha_{3}\left[\text{tr} \left(\Curl  \p\right) \right]^2 \right)\,,
\end{align}
with scalar weighting parameters $\alpha_{1},\alpha_{2},\alpha_{3}\geq0$.  Since, in this paper, the curvature energy does not play a major role we will mostly just use $\lVert \Curl\p\rVert^{2} $, corresponding to $\alpha_{1},\alpha_{2}=1$ and $\alpha_{3}=\frac{1}{3}$. . 

\subsubsection{Micro-inertia density \label{microinertia}}

The dynamical formulation of the proposed relaxed micromorphic model is obtained in the following way. We define a joint Hamiltonian and obtain the equations from the postulate of stationary action. In order to generalize the kinetic energy density to the anisotropic micromorphic framework, we need to introduce a micro-inertia density contribution of the type:
\begin{align}
\frac{1}{2}\langlenew \Jfourth \X \p_{,t},\p_{,t} \ranglenew.
\end{align}
Here $\Jfourth:\R^{3\times3}\rightarrow\R^{3\times3}$ is the $4^{th}$ order micro-inertia density tensor with, in general, 45 independent coefficients.  Eringen has added a conservation law for the micro-inertia density tensor $\Jfourth$, but in this work we assume a constant micro-inertia density tensor $\Jfourth$ as well as a constant mass density $\rho$. We assume throughout this paper that $\Jfourth$ is positive definite, i.e.:
\begin{align}
\exists \, c^{+} >0: \, \forall X \in :\R^{3\times3}:\quad \langlenew\Jfourth\X X,X\ranglenew_{\R^{3\times3}} \geq c^{+} \|X \|^2_{\R^{3\times3}}.
\end{align}
Considering dimensional consistency, we can always write the micro-inertia density tensor $\Jfourth$  as:
\begin{align}
\Jfourth=\rho\, \widehat{L}_{c}^2\,\Jfourth_{0},
\end{align}
where $\Jfourth_{0}:\R^{3\times3}\rightarrow\R^{3\times3}$ is dimensionless. Here, $\rho>0$ is the mean mass density $[\rho]=kg/m^3$ and $\widehat{L}_{c}\geq0$ is another characteristic length $[\widehat{L}_{c}]=m$. We also propose a split of this micro-inertia density, similar {to that adopted for the other elastic tensors} like:
\begin{align}
\frac{1}{2}\langlenew \Jfourth\X \p_{,t},\p_{,t}\ranglenew=\,\frac{1}{2}\langlenew \Jsym_{e}\X \sym \p_{,t},\sym \p_{,t}\ranglenew+\frac{1}{2}\langlenew \Jso_{c}\X \skew \p_{,t},\skew \p_{,t}\ranglenew .
\end{align}
Here, $\Jsym_{e}:\Sym(3)\rightarrow\Sym(3)$ maps symmetric tensors into symmetric tensors while $\Jso_{c}:\so(3)\rightarrow\so(3)$ maps skew-symmetric tensors to skew-symmetric tensors. We assume then that both $\Jsym_{e}$ and $\Jso_{c}$ are positive definite.

In the isotropic case, the micro-inertia density tensor $\Jfourth_{0}$ can be represented by three dimensionless parameters $\eta_{1},\,\eta_{2},\,\eta_{3}>0$ such that:
\begin{align}
\frac{1}{2}\langlenew \Jfourth\X \p_{,t},\p_{,t}\ranglenew=\,\frac{\rho \widehat{L}_{c}^2}{2} \left (\eta_{1} \left\|\text{dev}\sym \p_{,t} \right\|^{2}+ \eta_{2} \left\|\skew \p_{,t} \right\|^{2}+\eta_{3} \left(\text{tr} \left(\p_{,t} \right)\right)^{2}\right) .
\end{align}

\subsubsection{Linear elasticity as upper energetic limit for the relaxed micromorphic model - statics}

The relaxed micromorphic model admits linear elasticity as an upper energetic limit for any characteristic length scale $L_{c}>0$. This can be seen by noticing that an admissible field for the micro-distortion $\p$ is always $\p=\nablau$. Then, a standard minimization argument shows
\footnote{{The strict equality in \eqref{Ineq} is trivial considering that replacing $\p=\nablau$ on the left hand side and recalling that $\Curl\,\nabla \vartheta=0$. On the other hand, the inequality can be justified thinking that a solution $(u^*,\p^*)$ of the relaxed micromorphic problem is a minimizer, in the sense that $W(u^*,\p^*)\leq W(u,\p)$ for any admissible field $(u,\p)$. Hence, taking a generic field $\p=\nablau$ (which is of course admissible) justifies the equation \eqref{Ineq}.  }}:
\begin{align}
\underset{(u,\p)}{\mathrm{min}}&\bigg\{\int_{\Omega}\frac{1}{2}\langlenew\Ce \X \sym\left(\nablau-\p\right),\sym\left(\nablau-\p\right)\ranglenew_{\R^{3\times3}}+\frac{1}{2}\langlenew\Ch \X\sym \p,\sym \p\ranglenew_{\R^{3\times3}} \label{Ineq}\\
& \quad 
+\frac{1}{2}\langlenew\Cc \X \skew\left(\nablau-\p\right),\skew\left(\nablau-\p\right)\ranglenew_{\R^{3\times3}}
+\frac{\mLc}{2} \langlenew \Lf_{\text{aniso}} \X \, \Curl \p,\Curl \p\ranglenew_{\R^{3\times3}}\,dx \bigg\}\nonumber \\&\qquad \quad
\leq \int_{\Omega} \frac{1}{2} \langlenew \Ch\X\sym\nablau,\sym\nablau\ranglenew_{\R^{3\times3}}\, dx
\,.
\nonumber 
\end{align}
Thus, we see that the relaxed model is always energetically \textbf{weaker than a linear elastic comparison material} with elastic stiffness $\Ch$ for any given stiffness $\Ce$. This, again, is in contrast to the standard Mindlin-Eringen format which will, in general, generate arbitrary stiffer response as   $L_{c}\rightarrow\infty$ and $\Ce\rightarrow\infty$ simultaneously.

\subsection{Energy formulations and equilibrium equations for various symmetries}

Gathering our findings, we propose the following representation of the energy for the relaxed anisotropic centro-symmetric model, which has maximally 21+21+6+21+6=75 independent coefficients:
\begin{align}
W= & \underbrace{\frac{1}{2}\langlenew\Ce \X \sym\left(\nablau-\p\right),\sym\left(\nablau-\p\right)\ranglenew_{\R^{3\times3}}}_{\mathrm{{\textstyle anisotropic\ elastic-energy}}}+\frac{1}{2}\underbrace{\langlenew\Ch \X\sym \p,\sym \p\ranglenew_{\R^{3\times3}}}_{\mathrm{\textstyle micro-self-energy}}\\
& \ 
+\underbrace{\frac{1}{2}\langlenew\Cc \X \skew\left(\nablau-\p\right),\skew\left(\nablau-\p\right)\ranglenew_{\R^{3\times3}}}_{\begin{array}{c}
	\mathrm{\textstyle invariant\ local\ anisotropic} \\
	\mathrm{\textstyle rotational\ elastic\ coupling\ }	
	\end{array}
}\nonumber\\&\ 
+\underbrace{\frac{\mLc}{2}\big[\langlenew\Lsym_{e} \X \sym \Curl \p,\sym \Curl \p \ranglenew_{\R^{3\times3}} + \langlenew \Lsym_{c} \X \skew \Curl \p,\skew \Curl \p \ranglenew_{\R^{3\times3}}  \big]}_{\mathrm{\textstyle curvature}}\,.
\nonumber 
\end{align}
{This constitutive expression of the strain energy density for the relaxed micromorphic model is the most general one that can be provided in the anisotropic and centrosymmetric framework and already it provides a drastic reduction of the constitutive coefficients with respect to the standard Mindlin-Eringen model (75 coefficients against the 498 of Mindlin-Eringen). With a look towards immediate applications, it can be considered that non-local effects can be considered, in a first instance, to be isotropic, so that the curvature coefficients reduce from 21+6=27, to at most 2. We end up with a fully anisotropic model which features at most 51 parameters for describing:}
{
\begin{itemize}
	\item the full anisotropy at the microstructural level.
	\item the full anisotropy at the macroscopic level.
	\item the possibility of describing non-localities through the introduction of suitable characteristic lengths.
\end{itemize}}
{Of course, given particular metamaterials with particular symmetries, this number of parameters can be further reduced.}

{For example, the fully} isotropic case requires to determine ($\Ce\sim2,\,\Ch\sim2,\Cc\sim1,\,\Lsym_{e}\sim2,\,\Lsym_{c}\sim1$) altogether 8 constitutive coefficients of which the rotational coupling coefficient $\mc$ can be set to zero to enforce \textbf{symmetric} elastic stresses $\sigma$. As seen before, Eringen's formulation has 18 coefficients and Koh's \cite{koh1970special} micro-isotropic model has still 10.\footnote{Note that establishing positive-definiteness of the energy is now an easy matter as compared to \cite{smith1968inequalities}: we only need to require positive definiteness of the occurring standard 4$^{th}$ order tensors $\Ce,\Ch,\Cc,\Lsym_{e},\Lsym_{c}$.} 
{This simplified framework allowing to describe the full micro-macro anisotropy and the presence of non-localities via the introduction of ``only'' 51 parameters is of fundamental importance to proceed towards an enlightened characterization of the actual metamaterial.} 

Considering 
\begin{align}
\frac{\rho\, \widehat{L}_{c}^2}{2}\langlenew \Jfourth_{0} \X \p_{,t},\p_{,t} \ranglenew,
\end{align}
as the micro-inertia term, the dynamical equilibrium equations for the anisotropic relaxed micromorphic model take the compact format:
\begin{align}
\rho\, u_{,tt}=&\,\Div\left[\Ce \X \sym\left(\nablau-\p\right)+\Cc \X \skew\left(\nablau-\p\right)\right] , \nonumber \\
\rho\,\widehat{L}_c^2 \,\Jfourth_{0} \X  \p_{,tt}=&\,\Ce \X \sym\left(\nablau-\p\right) +\Cc \X \skew\left(\nablau-\p\right)-\Ch \X\sym \p \\&\ -\mLc  \Curl \left( \Lsym_{e}\X \sym \Curl\p +\Lsym_{c} \X \skew \Curl \p\right).\nonumber 
\end{align}
If we consider the isotropic case and the simplest curvature form, it is possible to reduce the relaxed representation to {(see  \cite{madeo2014band,madeo2015wave,neff2014unifying,madeo2016first,neff2017real,madeo2016complete})}:
\begin{align}
W=&\underbrace{\me\lVert \sym\left(\nablau-\p\right)\rVert ^{2}+\frac{\lle}{2}\mathrm{tr}\left(\sym\left(\nablau-\p\right)\right)^{2}}_{\mathrm{{\textstyle isotropic\ elastic-energy}}}+\underbrace{\mh\lVert \sym \p\rVert ^{2}+\frac{\lh}{2}\,\left(\mathrm{tr}\left(\sym \p\right)\right)^{2}}_{\mathrm{{\textstyle micro-self-energy}}}\label{eq:Ener-2}\\
& \quad 
+\hspace{-0.5cm}\underbrace{\mc\lVert \skew\left(\nablau-\p\right)\rVert ^{2}}_{\begin{array}{c}
	\mathrm{\textstyle invariant\ local\ isotropic} \\
	\mathrm{\textstyle rotational\   elastic\ coupling\ }	
	\end{array}
}\hspace{-0.5cm} 
+\underbrace{\frac{\mLc}{2} \lVert \Curl \p\rVert^2}_{\mathrm{\textstyle isotropic\ curvature}}\,,
\nonumber  
\end{align}
and the isotropic format of the micro-inertia becomes:
\begin{align}
\frac{1}{2}\langlenew \Jfourth\X \p_{,t},\p_{,t}\ranglenew=\,\frac{\rho \widehat{L}_{c}^2}{2} \left (\eta_{1} \left\|\text{dev}\sym \p_{,t} \right\|^{2}+ \eta_{2} \left\|\skew \p_{,t} \right\|^{2}+\eta_{3} \left(\text{tr} \left(\p_{,t} \right)\right)^{2}\right) .
\end{align}
{Hence,} the dynamical equilibrium equations for the isotropic relaxed micromorphic model take the form:
\begin{align}
\rho\, u_{,tt}=&\,\Div\left[\Ce \X \sym\left(\nablau-\p\right)+\Cc \X \skew\left(\nablau-\p\right)\right] , \\
\nonumber
\\
	\eta_{1}\, \rho\,\widehat{L}_c^2 \dev\,\sym \left[\p_{,tt}\right]=&\dev\,\sym\left[\Ce \X \sym\left(\nablau-\p\right)-\Ch \X\sym \p-\mLc\Curl \Curl \p \right]\,,\nonumber\\\nonumber\\
	\eta_{2}\, \rho\,\widehat{L}_c^2 \,\skew\left[\p_{,tt}\right]=&\,\Cc \X \skew\left(\nablau-\p\right)-\mLc \skew \Curl \Curl \p \,,\nonumber\\\nonumber\\
	\eta_{3}\, \rho\,\widehat{L}_c^2 \tr\left[\p_{,tt}\right]=&\,\mathrm{tr}\,\left[\Ce \X \sym\left(\nablau-\p\right)-\Ch \X\sym \p -\mLc\Curl \Curl \p \right]\,.\nonumber
\end{align}
{For more information about the dynamics of the relaxed micromorphic model see \cite{madeo2016reflection,neff2017real}.}

\subsection{Some considerations on the macroscopic consistency condition in the isotropic case}
{In this section, we want to recall some results concerning the \textbf{macroscopic consistency condition} for the relaxed micromorphic model in the isotropic case \cite{neff2004material,neff2007geometrically}.} 

{Although such condition has already been derived in \cite{neff2004material,neff2007geometrically} and even if it is only valid for the particular case of isotropy, we want to underline the idea which is behind such condition. Indeed, it is of fundamental importance to catch the power that the introduced homogenization formulas may have for an effective application of the relaxed micromorphic model. In section \ref{Macroscopic}, we will present a generalization of such homogenization formulas to the fully anisotropic framework so opening the way for the effective mechanical characterization of a huge class of mechanical metamaterials.}

{The main idea, which is behind the determination of our homogenized formulas, is that of considering a very large sample of a given microstructured material. This sample must be large enough that the effect of the microstructure on the macroscopic behavior of the sample can be considered to be negligible.}

{Under this hypothesis, we can introduce} a macroscopic elasticity tensor $\C:\Sym(3)\rightarrow\Sym(3)$ which best fits the macroscopic behavior {of the sample and we can suppose that the material behavior can be} described by classical linear elasticity with energy:
\begin{align}
W=\, & \frac{1}{2}\,\langlenew\C \X\sym\nablau ,\sym\nablau \ranglenew\label{eq:WClass}.
\end{align}
The corresponding classical symmetric Cauchy stress {is clearly defined as}:
\begin{align}
\sigma(\sym\nablau) =\, & \C \X\sym\nablau .\label{eq:SigmaClass}
\end{align}
For very large sample sizes, however, a scaling argument shows easily
that the \textbf{relative} characteristic length scale $L_{c}$ of
the micromorphic model must vanish. Therefore, we have a way of comparing
the classical formulation \eqref{eq:WClass} to {the relaxed micromorphic formulation} \eqref{eq:Ener3}
and to offer an \textbf{a priori relation} between $\Ce$, $\Ch$
on the one hand and $\C$ on the other.

In the isotropic case, this has been already done in \cite{neff2004material,neff2007geometrically} with the \textbf{isotropic macroscopic consistency conditions}:
\begin{align}
\left(2\mm+3\lm\right)= &\, \frac{\left(2\me+3\lle\right)\left(2\mh+3\lh\right)}{\left(2\me+3\lle\right)+\left(2\mh+3\lh\right)},\label{eq:IsotropicRel-2}\\\nonumber\\\nonumber\mm=&\,\frac{\me\,\mh}{\me+\mh}=\me\,\left(\me+\mh \right)^{-1}\,\mh.
\end{align}
Or, analogously:
\begin{align}
\left(2\me+3\lle\right)= &\, \frac{\left(2\mm+3\lm\right)\left(2\mh+3\lh\right)}{\left(2\mh+3\lh\right)-\left(2\mm+3\lm\right)},\label{eq:IsotropicRel}\\\nonumber\\\nonumber \me=&\,\frac{\mm\,\mh}{\mh-\mm}=\mm\,\left(\mh-\mm\right)^{-1}\,\mh.
\end{align}
Note that these formulas determine $\mm$ and $\km$ (the elastic bulk
modulus $\km=\frac{2\mm+3\lm}{3}$) to be one half of the harmonic
mean of $\me$, $\mh$, and $\ke$, $\kh$ respectively.\begin{framed}
	 As a matter of fact, the \textbf{harmonic mean} $\mathcal{H} \left(\me,\mh\right)$ defined for real numbers  is:
\begin{align}
\mathcal{H} \left(\me,\mh\right)=\left[\frac{1}{2}\left(\frac{1}{\me}+\frac{1}{\mh}\right) \right]^{-1}=\frac{2\, \me\,\mh}{\me+\mh}\,.
\end{align}
\end{framed}
In the isotropic case, upon inspection of formula \eqref{eq:IsotropicRel},
we see that the ``macroscopic'' elastic response, embodied by $\mm$
and $\lm$, cannot be equal or stiffer than the microscopic
response, embodied by $\mh$ and $\lh$. This is certainly physically sound and expresses in short that \textbf{ ``smaller is stiffer''}. Moreover, $\mh=\mm$ is tantamount to ``micro = macro'' and formally equivalent to $\me\rightarrow\infty$.

{The fundamental importance of formulas \eqref{eq:IsotropicRel} and \eqref{eq:IsotropicRel} has already been proven in \cite{madeo2016modeling}, where it is shown that the macroscopic stiffnesses provide the slopes of the acoustic curves for band-gap metamaterials. This will be even clearer in further applications where static test will be conceived to evaluate ``a priori'' $\lm$ and $\mm$.}

\section{Mandel-Voigt vector notation\label{sec:Mandel-Voigt-vector-notation}}
{In this section, we consider an equivalent formulation of the relaxed micromorphic model obtained by using the Mandel-Voig vector notation for the macro strain $\sim \nablau$ as well as for the micro strain $\sym\p$.

This means that the second order tensors $\sym\nablau$ and $\sym\p$ are replaced by the vectors $\varepsilon$ and $\beta$, in which the components of the original tensors are sorted column-wise by respecting a given order which is chosen ``a priori''.

As it will be shown in subsection \ref{VoigtSym}, the use of such vector notation allows to represent the fourth order tensors $\Ce$ and $\Ch$ in $\R^{3}$ as second order tensors $\Cte$ and $\Cth$ in $\R^{6}$.

This representation is more suitable if one wants to specify the anisotropy classes of $\Cte$ and $\Cth$ in a format that is easily found in the literature.

For completeness, also the coupling fourth order tensor $\Cc$ can be casted in the form of a second order tensor  $\Ctc$ in $\R^{3}$ by suitably arranging the non-vanishing components of the skew-symmetric second order tensor $\skew (\nablau-\p)$ in a vector $\gamma \in \R^3$. As it will be shown in subsection \ref{VoigtSkew}, also the anisotropy classes of tensors of the type of $\Ctc$ are easily found in the literature.
}

We consider the general anisotropic expression for the relaxed micromorphic
model (see equation \eqref{eq:Ener3}), given in index notation as:
\begin{align}
W=\, & \frac{1}{2}\left(\Ce\right)_{ijkl}\left(\sym\left(\nablau-\p\right)\right)_{ij}\left(\sym\left(\nablau-\p\right)\right)_{kl}+\frac{1}{2}\left(\Cc\right)_{ijkl}\left(\skew\left(\nablau-\p\right)\right)_{ij}\left(\skew\left(\nablau-\p\right)\right)_{kl}\nonumber\\
& \ +\frac{1}{2}\left(\Ch\right)_{ijkl}\left(\sym \p\right)_{ij}\left(\sym \p\right)_{kl}+\frac{\mLc}{2} \left( \p_{ia,b}\,\epsilon_{jab}\right)\left( \p_{ic,d}\,\epsilon_{jcd}\right) \,,
\label{eq:Energy}
\end{align}
where $\epsilon$ is the Levi-Civita tensor. {We recall that}  $\Ce$, $\Ch:\Sym(3)\rightarrow\Sym(3)$
have at most 21 independent constants, while $\Cc:\so(3)\rightarrow\so(3)$
has at most 6 independent constants. 

\subsection{{Determination of the symmetric second order tensors $\Cte$ and $\Cth$ in terms of $\Ce$ and $\Ch$}\label{VoigtSym}}

We now consider a linear mapping $\mathfrak{M}_{\alpha ij}:\Sym(3)\rightarrow\R^{6}$
(as done in \cite{voigt1889ueber,voigt1908lehrbuch,mandel1962plastic}) such that the independent components
of $\left(\sym\nablau \right)_{ij}$
are isomorphically mapped in a corresponding vector $\varepsilon_{\alpha }$
such as:
\begin{align}
\varepsilon_{\alpha }=\, & \mathfrak{M}_{\alpha ij}\left(\sym\nablau \right)_{ij}.\label{eq:eps}
\end{align}
And in the same fashion we have:
\begin{align}
\beta_{\alpha }= & \,\mathfrak{M}_{\alpha ij}\left(\sym \p\right)_{ij}.\label{eq:alfa}
\end{align}
Here and in the following, Latin subscripts range in $\{1,2,3\}$ while Greek subscripts vary in $\{1,2,3,4,5,6\}$. {Following Mandel and Voigt, we set:}
\begin{align}
\beta= & \left(\begin{array}{c}
\left(\sym \p\right)_{11}\\
\left(\sym \p\right)_{22}\\
\left(\sym \p\right)_{33}\\
c\left(\sym \p\right)_{23}\\
c\left(\sym \p\right)_{13}\\
c\left(\sym \p\right)_{12}
\end{array}\right),\qquad\qquad\varepsilon=\left(\begin{array}{c}
\left(\sym\nablau \right)_{11}\\
\left(\sym\nablau \right)_{22}\\
\left(\sym\nablau \right)_{33}\\
c\left(\sym\nablau \right)_{23}\\
c\left(\sym\nablau \right)_{13}\\
c\left(\sym\nablau \right)_{12}
\end{array}\right),\label{eq:alfa-1-1}
\end{align}
where the coefficient c depends on the notation used (2 for Voigt notation \cite{voigt1889ueber,voigt1908lehrbuch},
$\sqrt{2}$ for Mandel notation \cite{mandel1962plastic}) {and this defines the mapping $\mathfrak{M}$}. 

The components of the defined mapping $\mathfrak{M}_{\alpha ij}$ can be represented as $3\times3$ matrices once fixing the index $\alpha$, as:
\begin{align}
\mathfrak{M}_{1ij}= & \left(\begin{array}{ccc}
1 & 0 & 0\\
0 & 0 & 0\\
0 & 0 & 0
\end{array}\right),\qquad\,\ \mathfrak{M}_{2ij}=\left(\begin{array}{ccc}
0 & 0 & 0\\
0 & 1 & 0\\
0 & 0 & 0
\end{array}\right),\qquad\ \ \mathfrak{M}_{3ij}=\left(\begin{array}{ccc}
0 & 0 & 0\\
0 & 0 & 0\\
0 & 0 & 1
\end{array}\right),\nonumber \\
\\
\mathfrak{M}_{4ij}= & \left(\begin{array}{ccc}
0 & 0 & 0\\
0 & 0 & \frac{c}{2}\\
0 & \frac{c}{2} & 0
\end{array}\right),\qquad\ \mathfrak{M}_{5ij}=\left(\begin{array}{ccc}
0 & 0 & \frac{c}{2}\\
0 & 0 & 0\\
\frac{c}{2} & 0 & 0
\end{array}\right),\qquad\ \mathfrak{M}_{6ij}=\left(\begin{array}{ccc}
0 & \frac{c}{2} & 0\\
\frac{c}{2} & 0 & 0\\
0 & 0 & 0
\end{array}\right).\nonumber 
\end{align}
We define the inverse operator $\mathfrak{M}_{ij\alpha}^{-1}:\R^{6}\rightarrow\Sym(3)$
such that:
\begin{align}
\left(\sym\nablau \right)_{ij}=\, & \mathfrak{M}_{ij\alpha}^{-1}\ \varepsilon_{\alpha},\qquad\qquad\left(\sym \p\right)_{ij}=\,\mathfrak{M}_{ij\alpha}^{-1}\ \beta_{\alpha},\label{eq:eps-inv}
\end{align}
and such that the following property:
\begin{align}
\mathfrak{M}_{\alpha ij}\,\mathfrak{M}_{ij\beta}^{-1}= & \,\widetilde{\delta}_{\alpha \beta},
\end{align}
where $\widetilde{\delta}$ is the Kronecker $\delta$ in $\R^{6}\times\R^{6}$, holds. It is possible to show that the components of the inverse operator are:
\begin{align}
\mathfrak{M}_{ij1}^{-1}= & \left(\begin{array}{ccc}
1 & 0 & 0\\
0 & 0 & 0\\
0 & 0 & 0
\end{array}\right),\qquad\ \ \mathfrak{M}_{ij2}^{-1}=\left(\begin{array}{ccc}
0 & 0 & 0\\
0 & 1 & 0\\
0 & 0 & 0
\end{array}\right),\qquad\ \ \mathfrak{M}_{ij3}^{-1}=\left(\begin{array}{ccc}
0 & 0 & 0\\
0 & 0 & 0\\
0 & 0 & 1
\end{array}\right),\nonumber \\
\\
\mathfrak{M}_{ij4}^{-1}= & \left(\begin{array}{ccc}
0 & 0 & 0\\
0 & 0 & \frac{1}{c}\\
0 & \frac{1}{c} & 0
\end{array}\right),\qquad\ \mathfrak{M}_{ij5}=\left(\begin{array}{ccc}
0 & 0 & \frac{1}{c}\\
0 & 0 & 0\\
\frac{1}{c} & 0 & 0
\end{array}\right),\qquad\ \mathfrak{M}_{ij6}^{-1}=\left(\begin{array}{ccc}
0 & \frac{1}{c} & 0\\
\frac{1}{c} & 0 & 0\\
0 & 0 & 0
\end{array}\right).\nonumber 
\end{align}
The mapping $\mathfrak{M}$ has zeros everywhere except in the components $\{111,222,333,423,513,612\}$. Therefore, we can express it compactly as:
\begin{align}
\mathfrak{M}_{\alpha ij}=\ &\widetilde{\delta}_{\alpha 1}\delta_{i1}\delta_{j1}+\widetilde{\delta}_{\alpha 2}\delta_{i2}\delta_{j2}+\widetilde{\delta}_{\alpha 3}\delta_{i3}\delta_{j3}+\frac{c}{2}\left(\widetilde{\delta}_{\alpha 4}\left(\delta_{i2}\delta_{j3}+\delta_{i3}\delta_{j2}\right)+\widetilde{\delta}_{\alpha 5}\left(\delta_{i1}\delta_{j3}+\delta_{i3}\delta_{j1}\right)\right) \label{mapping}\\
 & \ +\frac{c}{2}\,\widetilde{\delta}_{\alpha 6}\left(\delta_{i1}\delta_{j2}+\delta_{i2}\delta_{j1}\right).\nonumber 
\end{align}
Analogously for the inverse $\mathfrak{M}^{-1}$:
\begin{align}
\mathfrak{M}_{ij\alpha}^{-1}= \ &\widetilde{\delta}_{\alpha1}\delta_{i1}\delta_{j1}+\widetilde{\delta}_{\alpha2}\delta_{i2}\delta_{j2}+\widetilde{\delta}_{\alpha3}\delta_{i3}\delta_{j3}+\frac{1}{c}\left(\widetilde{\delta}_{\alpha4}\left(\delta_{i2}\delta_{j3}+\delta_{i3}\delta_{j2}\right)+\widetilde{\delta}_{\alpha5}\left(\delta_{i1}\delta_{j3}+\delta_{i3}\delta_{j1}\right)\right)\label{eq:MappInv}\\
 & \ +\frac{1}{c}\,\widetilde{\delta}_{\alpha6}\left(\delta_{i1}\delta_{j2}+\delta_{i2}\delta_{j1}\right).\nonumber 
\end{align}
It can be checked that applying the linear mapping \eqref{mapping} to a symmetric second order tensor $s_{ij}$, the result is a vector in $\R^{6}$ whose first 3 components are the elements $s_{11},s_{22}$ and $s_{33}$, while its last 3 components are $c\,s_{23}$, $c\,s_{13}$ and $c\,s_{12}$, respectively. This is consistent with the classical notation of equation \eqref{eq:alfa-1-1}. 

Now, if we consider a quadratic energy in $\varepsilon$ - $\beta$, {recalling equation \eqref{eq:eps} and \eqref{eq:alfa}, we can write it as:}
\begin{align}
 & \frac{1}{2}\left(\Cte\right)_{\alpha \beta}\left(\varepsilon_{\alpha}-\beta_{\alpha}\right)\left(\varepsilon_{\beta}-\beta_{\beta}\right)=\frac{1}{2}\left(\Cte\right)_{\alpha \beta}\mathfrak{M}_{\alpha ij}\mathfrak{\,M}_{\beta kl}\left(\sym\left(\nablau-\p\right)\right)_{ij}\left(\sym\left(\nablau-\p\right)\right)_{kl},\label{eq:EnerVect}
\end{align}
where, $\Cte:\R^{6}\rightarrow\R^{6}$
is a general second order symmetric tensor on $\R^{6\times 6}$(matrix), with 21 independent
coefficients.

Comparing equation (\ref{eq:EnerVect}) with the corresponding part of (\ref{eq:Energy}),
i.e.: 
\begin{align}
&\frac{1}{2}\left(\Ce\right)_{ijkl}\left(\sym\left(\nablau-\p\right)\right)_{ij}\left(\sym\left(\nablau-\p\right)\right)_{kl}= \label{eq:Cu-1}\\&\qquad \frac{1}{2}\left(\Cte\right)_{\alpha \beta}\mathfrak{M}_{\alpha ij}\mathfrak{\,M}_{\beta kl}\left(\sym\left(\nablau-\p\right)\right)_{ij}\left(\sym\left(\nablau-\p\right)\right)_{kl},\nonumber
\end{align}
we must have:
\begin{empheq}[box=\widefbox]{align}
\left(\Ce\right)_{ijkl}= &\ \mathfrak{M}_{\alpha ij}\left(\Cte\right)_{\alpha \beta}\mathfrak{M}_{\beta kl}\,.\label{eq:Cu}
\end{empheq}
For what follows, it is useful to remark that:
\begin{empheq}[box=\widefbox]{align}
\left(\Ce\right)_{ijkl}^{-1}= & \ \mathfrak{M}_{ ij\alpha}^{-1} \left(\Cte\right)_{\alpha \beta}^{-1} \mathfrak{M}_{ kl\beta}^{-1}\,\label{eq:Cu2}.
\end{empheq}
This last relation is not trivial and is proven in the Appendix \ref{Dem}. 
 On the other hand, the converse relations read:
\begin{empheq}[box=\widefbox]{align}
\left(\Cte\right)_{\alpha \beta}=  \mathfrak{M}_{ij \alpha}^{-1} \left(\Ce\right)_{ijkl} \mathfrak{M}_{ kl \beta}^{-1}\,,\label{eq:CuInv}\qquad\qquad
\left(\Cte\right)_{\alpha \beta}^{-1}=\mathfrak{M}_{\alpha ij } \left(\Ce\right)_{ijkl}^{-1}  \mathfrak{M}_{\beta kl}
\,.
\end{empheq}
Using \eqref{eq:CuInv} and recalling expression \eqref{eq:MappInv} for the components of $\mathfrak{M}^{-1}$, it can be seen that the second order tensor $\Cte$ can be written as a function of the components of
the fourth order tensor $\Ce$:
\begin{align}
 & \Cte=\left(\begin{array}{cccccc}
 \vspace{0.1cm}
\left(\Ce\right)_{1111} & 
\left(\Ce\right)_{1122} & 
\left(\Ce\right)_{1133} & 
\frac{2}{c}\left(\Ce\right)_{1123} & 
\frac{2}{c}\left(\Ce\right)_{1113} &
 \frac{2}{c}\left(\Ce\right)_{1112}\\\vspace{0.1cm}
\left(\Ce\right)_{2211} & 
\left(\Ce\right)_{2222} & 
\left(\Ce\right)_{2233} & 
\frac{2}{c}\left(\Ce\right)_{2223} & 
\frac{2}{c}\left(\Ce\right)_{2213} & \frac{2}{c}\left(\Ce\right)_{2212}\\\vspace{0.1cm}
\left(\Ce\right)_{3311} & 
\left(\Ce\right)_{3322} & 
\left(\Ce\right)_{3333} & 
\frac{2}{c}\left(\Ce\right)_{3323} & 
\frac{2}{c}\left(\Ce\right)_{3313} & \frac{2}{c}\left(\Ce\right)_{3312}\\\vspace{0.1cm}
\frac{2}{c}\left(\Ce\right)_{2311} & \frac{2}{c}\left(\Ce\right)_{2322} & \frac{2}{c}\left(\Ce\right)_{2333} & \frac{4}{c^{2}}\left(\Ce\right)_{2323} & \frac{4}{c^{2}}\left(\Ce\right)_{2313} & \frac{4}{c^{2}}\left(\Ce\right)_{2312}\\\vspace{0.1cm}
\frac{2}{c}\left(\Ce\right)_{1311} & \frac{2}{c}\left(\Ce\right)_{1322} & \frac{2}{c}\left(\Ce\right)_{1333} & 
\frac{4}{c^{2}}\left(\Ce\right)_{1323} & \frac{4}{c^{2}}\left(\Ce\right)_{1313} & 
\frac{4}{c^{2}}\left(\Ce\right)_{1312}\\\vspace{0.1cm}
\frac{2}{c}\left(\Ce\right)_{1211} & \frac{2}{c}\left(\Ce\right)_{1222} &
\frac{2}{c}\left(\Ce\right)_{1233} & \frac{4}{c^{2}}\left(\Ce\right)_{1223} &
\frac{4}{c^{2}}\left(\Ce\right)_{1213} & 
\frac{4}{c^{2}}\left(\Ce\right)_{1212}
\end{array}\right)\,,\label{eq:Cu-2}
\end{align}
which is a symmetric $6\times6$ matrix due to the symmetries of $\Ce$ according to which
\begin{align}
\left(\Ce\right)_{ijkl}=\left(\Ce\right)_{klij}\,.
\end{align}
In the same fashion, we have the {relationships involving $\Ch$ and $\Cth$ as well as $\C$ and $\Ct$:}
\begin{empheq}[box=\widefbox]{align}
\left(\Ch\right)_{ijkl}= &\ \mathfrak{M}_{\alpha ij} \left(\Cth\right)_{\alpha \beta} \mathfrak{M}_{\beta kl},\qquad\qquad\left(\C\right)_{ijkl}=\mathfrak{M}_{\alpha ij} \left(\Ct\right)_{\alpha \beta} \mathfrak{M}_{\beta kl}\,,\label{eq:Cp}\\
\left(\Ch\right)^{-1}_{ijkl}= &\ \mathfrak{M}^{-1}_{ij \alpha} \left(\Cth\right)^{-1}_{\alpha \beta} \mathfrak{M}^{-1}_{kl \beta },\qquad\qquad\left(\C\right)^{-1}_{ijkl}=\mathfrak{M}^{-1}_{ij \alpha} \left(\Ct\right)^{-1}_{\alpha \beta} \mathfrak{M}^{-1}_{kl \beta}\,,\nonumber
\end{empheq}
and, conversely:
\begin{empheq}[box=\widefbox]{align}
\left(\Cth\right)_{\alpha \beta}= &\ \mathfrak{M}^{-1}_{ ij \alpha} \left(\Ch\right)_{ijkl} \mathfrak{M}^{-1}_{ kl \beta},\qquad\qquad\left(\Ct\right)_{\alpha \beta}=\mathfrak{M}^{-1}_{ij \alpha}\left(\C\right)_{ijkl} \mathfrak{M}^{-1}_{kl\beta},\\
\left(\Cth\right)^{-1}_{\alpha \beta}= &\ \mathfrak{M}_{ \alpha ij} \left(\Ch\right)^{-1}_{ijkl} \mathfrak{M}_{ \beta kl},\qquad\qquad\left(\Ct\right)^{-1}_{\alpha \beta}=\mathfrak{M}_{ \alpha ij} \left(\C\right)^{-1}_{ijkl} \mathfrak{M}_{\beta kl}.
\end{empheq}

\subsection{{Determination of the fourth order tensors $\Cc$  in terms of $\Ctc$} \label{VoigtSkew}}
{In this subsection, we extend the reasoning used in the previous subsection for the elastic tensors acting on symmetric strain measure to the elastic tensor $\Cc$ which instead acts on skew-symmetric strain measures and so provide the ``rotational coupling'' in the relaxed micromorphic model

To this aim, we} may always represent the $4^{th}$ order tensor $\Cc:\so(3)\rightarrow\so(3)$ acting on skew-symmetric matrices in its version acting on axial vectors only, i.e. we write:
\begin{align}
\langlenew \Cc \X \skew\left(X\right),\skew\left(X\right) \ranglenew_{\R^{3\times 3}} = \langlenew \Ctc\,\axl\left(\skew\left(X\right)\right),\axl\left(\skew\left(X\right)\right) \ranglenew_{\R^{3}},\label{eq:energyskew}
\end{align}
where $\Ctc:\R^{3}\rightarrow\R^{3}$ is a symmetric second order tensor (since it appears in a quadratic form) { and the operator $\axl$ defined in equation \eqref{axl}}. Therefore, $\Ctc$ has only 6 independent coefficients and so does $\Cc$. Given a second order tensor $X$, it can be verified that:
\begin{align}
\left\| \skew\left(X\right) \right\|^2_{\R^{3\times3}} =2\left\| \axl\left(\skew\left(X\right)\right) \right\| ^2_{\R^{3}}\,.
\end{align}
Before understanding the general anisotropic character of the coupling tensor $\Cc$, we recall the transformation behavior of the energy expression in the isotropic case. An energy defined on second order tensors is  isotropic if the transformation:
\begin{align}
X\rightarrow Q^T \x X \x Q\qquad \text{for}\quad Q \in \mathrm{SO}(3),\label{eq:transf}
\end{align}
does not affect the value of the energy. More precisely, we say that a local energy contribution acting on second order tensors is isotropic if
\begin{align}
W(X)=W(Q^T \x X \x Q).
\end{align}
Given a second order tensor which is subjected to the transformation \eqref{eq:transf}, it is clear that its skew-symmetric part transforms as follows:
\begin{align}
\skew\left(X\right)\rightarrow \skew\left(Q^T\x X \x Q\right)=Q^T \x \skew\left( X \right) \x Q\qquad \text{for} \ Q \in \mathrm{SO}(3)\, ,
\end{align}
and the corresponding axial vector of $\skew\left( X\right) $ satisfies the transformation law:
\begin{align}
\axl\left( \skew\left( X\right)\right) \rightarrow \axl\left( \skew\left( Q^T \x X \x Q\right)\right) &=\axl\left( Q^T \x \skew\left( X\right) \x Q \right)=Q \x \axl\left( \skew\left( X\right)\right),\label{eq:axial}
\end{align}
see \cite{munch2016rotational,munch2016modified}.
Based on these transformation laws, we may investigate the anisotropy of the rotational coupling with the representation in terms of { the second order tensor} $\Ctc$. Indeed, for the isotropy of an energy of the type $W(\skew X)$ we require the invariance:
\begin{align}
\forall\, Q \in \mathrm{SO}(3):\quad \langlenew \Cc \X \skew\left(X\right),\skew\left(X\right) \ranglenew_{\R^{3\times 3}} =\langlenew \Cc \X \skew\left(Q^T\x X\x Q\right),\skew\left(Q^T\x X\x Q\right) \ranglenew_{\R^{3\times 3}} ,
\end{align}
which, recalling \eqref{eq:energyskew} and \eqref{eq:axial} is also equivalent to:
\begin{align}
\forall\, Q \in \mathrm{SO}(3):\quad&\langlenew \Ctc \x \axl\left(\skew\left(X\right)\right),\axl\left(\skew\left(X\right)\right) \ranglenew_{\R^{3}}\nonumber\\&\quad=\langlenew \Ctc \x \axl\left(\skew\left(Q^T \x X \x Q\right)\right),\axl\left(\skew\left(Q^T \x X \x Q\right)\right) \ranglenew_{\R^{3}}\\&\quad
=\langlenew \Ctc \x Q \x \axl\left(\skew\left(X\right)\right) , Q \x \axl\left(\skew\left(X\right)\right) \ranglenew_{\R^{3}}\,.\nonumber
\end{align}
If we now set $\eta=\axl\left(\skew\left(X\right)\right)$, the latter is equivalent to:
\begin{align}
\forall\, Q \in \mathrm{SO}(3):\qquad \langlenew \Ctc \x \eta,\eta \ranglenew_{\R^{3}}=\langlenew \Ctc \x Q \x \eta,Q \x \eta \ranglenew_{\R^{3}}=\langlenew Q^T \x \Ctc \x Q \,\eta,\eta \ranglenew_{\R^{3}}  \label{eq:Quadratic}\,,
\end{align}
where the transformation laws for the $\axl$-operator given in \eqref{eq:axial} has been used. Since \eqref{eq:Quadratic} must hold for all vectors $\eta \in \R^3$ we obtain:
\begin{align}
\Ctc = Q^T \x \Ctc \x Q \qquad \forall\,Q \in \mathrm{SO}(3)\,.
\end{align}
Recalling that $Q \in \mathrm{SO}(3)$ implies $Q^{T}=Q^{-1}$, it can be inferred that this last equation is satisfied if and only if:
\begin{align}
\Ctc=\frac{\mc}{2}\, \mathds{1},\qquad \mc \geq 0,
\end{align}
which is the expression of $\Ctc$ for the isotropic case in which $\mc$ is called the \textbf{Cosserat couple modulus} \cite{cosserat1909theorie}.
Let us first state again that the relaxed micromorphic model is fully functional even without using $\Cc$ at all. However, our experience in the isotropic case, in which $\Cc$ reduces to the Cosserat couple modulus $\mc$, has shown that in order to describe complete frequency band gaps, one should take $\mc>0$. In the anisotropic case this would translate to requiring that $\Cc$ is positive definite.

Next, we discuss the different anisotropy classes for $\Cc$ which can be expressed more easily for $\Ctc$. In this case we discuss the solutions of:
\begin{align}
\Ctc = Q^T \x \Ctc \x Q \qquad \forall\,Q \in \mathcal{G}\text{\,-\,symmetry\ group\ of\ the\ material}.
\end{align}
In other words, the invariance condition is formally equivalent to that previously discussed (see equation \eqref{eq:transf}), but the difference stays in the set $\mathcal{G}$ in which the transformation matrix $Q$ lives. Depending on the symmetry properties of the group $\mathcal{G}$, we will be able to define different material classes.

It can be shown that in the respective cases of triclinic, monoclinic, orthorombic, tetragonal (coincides with transversely isotropy) and isotropy (equivalent to cubic symmetry), the tensor $\Ctc$ has the following forms (see \cite[p. 30]{lovett1989tensor} and \cite{chadwick2001new})\footnote{The most general representation of $\Ctc \in \text{Sym}^{+}(3)$ is $\Ctc =\text{dev}\,(\Ctc)+\frac{1}{3} \text{tr}(\Ctc)\mathds{1}$. However, this has nothing to do with isotropy: it is just a convenient representation for general symmetric $\Ctc$. We note in passing that $\dev \, \Ctc$ alone cannot be positive definite since $\tr \left(\dev  \,\Ctc\right)=0$, so there are positive and negative eigenvalues of $\dev \, \Ctc$.}:
\begin{align}
\Ctc^{\text{tric}}=&\left(\begin{array}{ccc}
\left(\Ctc\right)_{11} &\left(\Ctc\right)_{21}&\left(\Ctc\right)_{31}\\&\left(\Ctc\right)_{22}&\left(\Ctc\right)_{23} \\ \sym & & \left(\Ctc\right)_{33}
\end{array} \right),
\qquad  &&\Ctc^{\text{mono}}=\left(\begin{array}{ccc}
 \left(\Ctc\right)_{11} & 0 &\left(\Ctc\right)_{31}\\&\left(\Ctc\right)_{22}& 0 \\ \sym  & & \left(\Ctc\right)_{33}
\end{array} \right), 
 \nonumber\\ \Ctc^{\text{orth}}=&\left(\begin{array}{ccc}
 \left(\Ctc\right)_{11} & 0 & 0\\&\left(\Ctc\right)_{22} & 0 \\ \sym  & & \left(\Ctc\right)_{33}
\end{array} \right),
 \qquad
 &\Ctc^{\text{tetr}}=&\,\Ctc^{\text{trans}}=\left(\begin{array}{ccc}
 \left(\Ctc\right)_{11} & 0 & 0\\&\left(\Ctc\right)_{11} & 0 \\ \sym  & & \left(\Ctc\right)_{33}
\end{array} \right),\label{eq:CcSym} \\
&&\hspace{-1cm}\Ctc^{\text{iso}}=\,\Ctc^{\text{cubic}}=&\, \left(\Ctc\right)_{11}\left(\begin{array}{ccc}
	1 & 0 & 0 \\& 1 & 0 \\ \sym  & &1 \end{array} \right).\quad\qquad\nonumber
\end{align}

After considering the representation \eqref{eq:CcSym} we appreciate the fact that there is no difference between the cubic and isotropic rotational coupling. Both reduce $\Ctc$ to be a spherical tensor $
\Ctc=\frac{\mc}{2}\, \mathds{1},$ with $\mc \geq 0$. We believe that it is very difficult to make statements about the anisotropic rotational coupling, see the footnote \ref{footKroner}.

{Indeed, the first applications of the relaxed micromorphic model to real band-gap metamaterials show that an isotropic version of the tensor $\Ctc$ is sufficient to trigger band-gap behaviors. We provide in this paper the general framework to treat any possible degree of anisotropy for the rotational coupling. Nevertheless, if there is no evidence of the need of anisotropic rotational coupling based on experimental observations, an isotropic coupling given by the Cosserat couple modulus $\mc$ alone should always be preferred. Therefore, it is possible to consider a reduction of a given anisotropic rotational coupling to the isotropic case as analyzed in Appendix \ref{Coupling}.}

\section{The macroscopic limit of the relaxed model ($L_{c}\rightarrow 0$) - macroscopic consistency conditions}
{ In this section we provide one of the main findings of the present paper, namely a clear procedure for the determination of the macroscopic fourth order tensor $\C$ in terms of the microstructure-related $\Ce$ and $\Ch$. Thanks to our previous considerations, we are able to establish equivalent relationships between the second order tensors $\Ct$, $\Cte$ and $\Cth$. The results that we show in this section have the following advantages which allow us to expectedly proceed towards well-conceived applications on real metamaterials:
\begin{itemize}
	\item the consistency condition that we derive here relates the macro moduli in $\C$ to the micro moduli in $\Ce$ and $\Ch$. We claim that, given a specific metamaterial, the moduli in $\C$ can be determined on the basis of very simple mechanical tests. The idea is that of considering a specimen which is big enough that the effect of the microstructure can be considered to be negligible. Once the tensor $\C$ is known, then $\Ce$ and $\Ch$ can be directly related via the consistency condition that we present here. This drastically reduces the number of unknown coefficients that have to be determined, so providing an effective tool towards manageable applications.
\item the way towards application is made even easier by the introduction of the second order tensors $\Ct$, $\Cte$ and $\Cth$ whose form can be easily found in the literature once the class of anisotropy of the medium is fixed. 
\end{itemize} }

\subsection{Equilibrium equations}
 
The equilibrium equations of the anisotropic relaxed micromorphic model associated to the energy \eqref{eq:Ener3} read:
	\begin{align}
		\Div\left[\Ce \X \sym\left(\nablau-\p\right)+\Cc \X \skew\left(\nablau-\p\right)\right]  &=0,\label{eq:Equil} \\ \nonumber
		\\
		\Ce \X \sym\left(\nablau-\p\right) +\Cc \X \skew\left(\nablau-\p\right)\qquad\qquad\nonumber \\
		\qquad\ -\,\Ch\X \sym \p-\mLc\left(\Curl\Curl\p\right)&=0,\nonumber 
	\end{align}
or, in index notation:
\begin{align}
\left[\left(\Ce\right)_{ijkl}\left(\sym\left(\nablau-\p\right)\right)_{kl}+\left(\Cc\right)_{ijkl}\left(\skew\left(\nablau-\p\right)\right)_{kl}\right]_{,j}  &=0, \label{eq:Equil-2}\\
\nonumber\\
\left(\Ce\right)_{ijkl}\left(\sym\left(\nablau-\p\right)\right)_{kl} +\left(\Cc\right)_{ijkl}\left(\skew\left(\nablau-\p\right)\right)_{kl}\qquad\quad\nonumber \\
-\left(\Ch\right)_{ijkl}\left(\sym \p\right)_{kl} -\mLc\left(\p_{ik,jk}-\p_{ij,kk} \right) &=0.\nonumber 
\end{align}
We define the elastic stress tensor $\sigma(\nablau,\p)$ appearing in \eqref{eq:Equil}$_{1}$ as:
\begin{align}
\sigma(\nablau,\p):=\, & \Ce \X \sym\left(\nablau-\p\right)+\Cc \X \skew\left(\nablau-\p\right),\label{eq:SigmaMicro-1-1}
\end{align}
or, in index notation:
\begin{align}
\sigma_{ij}(\nablau,\p):= & \left(\Ce\right)_{ijkl}\left(\sym\left(\nablau-\p\right)\right)_{kl}+\left(\Cc\right)_{ijkl}\left(\skew\left(\nablau-\p\right)\right)_{kl}.\label{eq:SigmaMicro-1} 
\end{align}
Therefore, the equilibrium equation \eqref{eq:Equil}$_{1}$ can be compactly written as:
\begin{align}
\Div\left[\sigma(\nablau,\p)\right]=\, & 0.
\end{align}

\subsection{The general relaxed anisotropic case in the limit $L_{c}\rightarrow 0$ \label{Macroscopic}}

We will show that our relaxed micromorphic model defined by the energy \eqref{eq:Ener3}, or equivalently by the equations of motion \eqref{eq:Equil}, can be reduced to a sort of equivalent ``macroscopic model'' when letting $L_{c}\rightarrow 0$. Indeed, when $L_{c}=0$ equation \eqref{eq:Equil}$_{2}$ gives a direct relation between $\p$ and $\nablau$ which, when inserted in \eqref{eq:Equil}$_{1}$, allows to rewrite the energy in terms of $\nablau$.  Hence, we can introduce an equivalent macroscopic stress tensor $\sigma_{\mathrm{macro}}(\sym\nablau)$ which is the limit of  $\sigma(\nablau,\p)$ for $L_{c}\rightarrow 0$. In symbols:
\begin{align}
\sigma_{\mathrm{macro}}(\sym\nablau)=\lim_{L_{c}\rightarrow 0}\sigma(\nablau,\p)\,.
\end{align}
{ In the linear-elastic case }the tensor $\sigma_{\mathrm{macro}}(\sym\nablau)$ can be written as:
\begin{align}
\sigma_{\mathrm{macro}}(\sym\nablau)=\C\X\sym\nablau\,,
\end{align}
assuming that it is the Cauchy stress tensor of a classical first gradient continuum.

{ In view of applications,} considering very large samples of the anisotropic { medium is equivalent} to letting $L_{c}$, the characteristic length, tend to zero.
As a consequence of $L_{c}=0$, the second equilibrium equation in
(\ref{eq:Equil}) looses the $\Curl\Curl\p$-term
and turns into an algebraic side condition connecting $\p$
and $\nablau$ via:
\begin{align}
 & \Ce \X \sym\left(\nablau-\p\right)-\Ch \X\sym \p+\Cc \X \skew\left(\nablau-\p\right)=0,\label{eq:Equil2-2}
\end{align}
which, in index notation reads:
\begin{align}
 & \left(\Ce\right)_{ijkl}\left(\sym\left(\nablau-\p\right)\right)_{kl}-\left(\Ch\right)_{ijkl}\left(\sym \p\right)_{kl}+\left(\Cc\right)_{ijkl}\left(\skew\left(\nablau-\p\right)\right)_{kl}=0.\label{eq:Equil3}
\end{align}
Equation (\ref{eq:Equil2-2}) can be decoupled (by the assumed special mapping symmetry properties of the elasticity tensors, see equations \eqref{eq:Decom}) into two equations for the symmetric and skew-symmetric part, respectively, yielding:
\begin{align}
\Ce \X \sym\left(\nablau-\p\right)=\Ch \X\sym \p,\label{eq:Balance-1}\qquad \qquad
\Cc \X \skew\left(\nablau-\p\right)=\,0.
\end{align}
which, in index notation becomes:
\begin{align}
\left(\Ce\right)_{ijkl}\left(\sym\left(\nablau-\p\right)\right)_{kl}&=\left(\Ch\right)_{ijkl}\left(\sym\p\right)_{kl},\label{eq:Balance}\\ \nonumber
\left(\Cc\right)_{ijkl}\left(\skew\left(\nablau-\p\right)\right)_{kl}&=0.
\end{align}
This uncoupling is true since $\Ce$ and $\Ch$ map symmetric matrices to symmetric matrices and $\Cc:\so(3)\rightarrow\so(3)$, and then $\Cc \X \skew\left(\nablau-\p\right)$
is skew-symmetric by assumption. From the second equation in \eqref{eq:Balance-1}, we can easily derive that:
\begin{align}
\Cc\X\skew \nablau =\Cc\X \skew \left(\p\right). \label{eq:skeq}
\end{align}
{ On the other hand, solving} \eqref{eq:Balance-1}$_{1}$ for $\sym \p$
gives\footnote{We note here that the inverse of an elastic stiffness tensor, like $\left(\Ch+\Ce\right)$ has the same symmetry group structure as $\Ch+\Ce$ itself. This can be shown easily by directly looking at its definition { of} groups.}:
\begin{align}
\left(\Ch+\Ce\right) \X \sym \p= &\, \Ce \X \sym\nablau , \label{eq:SYMP} \\ 
\iff\quad\sym \p= & \left(\Ch+\Ce\right)^{-1}\X\left(\Ce \X \sym\nablau \right).\nonumber 
\end{align}
This is an identity between the micro-distortion $\p$ and the gradient of the displacement $\nablau$ which proves how, in the macroscopic limiting case, the model is transparent with respect to the micro-distortion{, i.e. only macroscopic deformations involving $\sym\nablau$ are allowed}.
We insert \eqref{eq:Balance-1}$_{1}$, \eqref{eq:skeq} and \eqref{eq:SYMP} into \eqref{eq:Equil}
and considering the uncoupling between symmetric and skew symmetric parts of the involved tensors, we get:
\begin{align}
\Div\left[\Ch \X\sym \p\right] & =0
\iff 
\Div\left[\Ch \X\left(\Ch+\Ce\right)^{-1} \X \, \Ce \X \sym\nablau \right]  =0.\label{eq:DivFinal}
\end{align}
On the other hand, the classical balance equation for the linear elastic macroscopic response is:
\begin{align}
\Div\left[\C \X\sym\nablau \right] & =0.\label{eq:macrbal}
\end{align}
Comparing the macroscopic balance equation \eqref{eq:macrbal} with the one derived from our relaxed model when letting $L_{c}=0$ (\eqref{eq:DivFinal}$_{1}$), we obtain the following \textbf{a priori relation} between the macroscopic elasticity tensor $\C$ and the microscopic tensor $\Ch$ as well as the mesoscopic (relative) elasticity tensor $\Ce$:
\begin{empheq}[box=\widefbox]{align} \C  :=\Ch \X\left(\Ch+\Ce\right)^{-1} \X\, \Ce\,, \label{eq:Relation}	
\end{empheq}
which is a generalization of \eqref{eq:IsotropicRel-2} when considering our anisotropic setting. From equation \eqref{eq:Relation} (see Appendix \ref{sec:Prop}), we get by simple inversion\footnote{{ It can be checked that, given fourth order invertible tensors $A$, $B$ and $C$, the following identity holds: $\left(A\cdot B\cdot C\right)^{-1}=C^{-1} \cdot B^{-1} \cdot A^{-1}$}}:
\begin{align}
\C^{-1} =\Ch^{-1} \X\left(\Ch+\Ce\right) \X\, \Ce^{-1}=\Ce^{-1}+\Ch^{-1}\,.\label{eq:CInv}
\end{align}
Therefore, we note, surprisingly at first glance, that $\C$ is the ``parallel sum'' of  $\Ce$ and $\Ch$ (the parallel sum of two tensors $A$ and $B$ is defined as  $\left(A^{-1}+B^{-1}\right)^{-1}$), that is equal to one half of the \textbf{harmonic mean operator} on positive definite symmetric matrices (see \cite[p. 103]{bhatia2009positive}), defined as:
\begin{align}
\mathcal{H} \left(\Ce ,\Ch \right):=
\left[\frac{1}{2} \left(\Ce^{-1}+\Ch^{-1}\right) \right]^{-1}=
2\, \Ch \X\left(\Ce+\Ch \right)^{-1} \X \, \Ce =2\, \C \,.
\end{align}
It is possible to obtain the inverse relation with algebraic operations. 
First, from equation \eqref{eq:CInv} it is immediate that:
\begin{align}
\Ce^{-1}=\C^{-1}-\Ch^{-1}\, ,
\end{align}
or equivalently:
\begin{align}
\Ce=&\left(\C^{-1}-\Ch^{-1} \right)^{-1}=\Ch \X \left[ \Ch^{-1} \X \left(\C^{-1}-\Ch^{-1} \right)^{-1}\X\, \C^{-1} \right] \X\, \C\,.
\end{align}
Considering that $A^{-1} \cdot B^{-1} \cdot C^{-1}=\left(C\cdot B\cdot A\right)^{-1}$, we obtain:
\begin{align}
\Ce=&\, \Ch \X\left[\C \X \left(\C^{-1}-\Ch^{-1} \right) \X \, \Ch \right]^{-1} \X\, \C \\ \nonumber =& \,\Ch \X\left[\Ch-\C\right]^{-1} \X \,\C \,.
\end{align}
So finally, we have the further compact relation:
\begin{empheq}[box=\widefbox]{align}
\Ce=\Ch \X\left(\Ch-\C\right)^{-1} \X\, \C\,.
\end{empheq}
Note that these results are true without assuming that the tensors $\Ch$, $\Ce$ and $\C$
commute (and, in fact, they do not).

\subsection{Particularization for specific { anisotropy classes}}

In order to show how equation \eqref{eq:Relation} particularizes  { for anisotropy classes}, we use
the vectorial notation defined in section (\ref{sec:Mandel-Voigt-vector-notation}). In particular, by \eqref{eq:Relation},   \eqref{eq:Cu} and \eqref{eq:Cu2}, we can rewrite equation \ref{eq:Relation}:  
 \begin{align}
  \left(\Ct\right)_{\alpha \beta} \mathfrak{M}_{\alpha ij}\ \mathfrak{M}_{\beta kl}  =&\left(\Cth\right)_{\alpha \gamma} \mathfrak{M}_{\alpha ij}\, \mathfrak{M}_{\gamma mn}   \left(\Cth+\Cte\right)^{-1}_{\delta \epsilon} \mathfrak{M}^{-1}_{mn \delta}\, \mathfrak{M}^{-1}_{pq\epsilon}  \left(\Cte\right)_{\zeta \beta} \mathfrak{M}_{\zeta pq}\ \mathfrak{M}_{\beta kl}\nonumber\\
  =& \left(\Cth\right)_{\alpha \gamma} \widetilde{\delta}_{\gamma \delta}    \widetilde{\delta}_{\epsilon \zeta}  \left(\Cth+\Cte\right)^{-1}_{\delta \epsilon}  \left(\Cte\right)_{\zeta \beta} \mathfrak{M}_{\alpha ij}\, \mathfrak{M}_{\beta kl}\\
    =& \left(\Cth\right)_{\alpha \gamma} \left(\Cth+\Cte\right)^{-1}_{\gamma \zeta} \left(\Cte\right)_{\zeta \beta} \mathfrak{M}_{\alpha ij}\, \mathfrak{M}_{\beta kl}\,.\nonumber
 \end{align} 
 From this last equation we easily notice that:
\begin{empheq}[box=\widefbox]{align}
 \Ct  =\Cth \x \left(\Cth+\Cte\right)^{-1}  \x \Cte\,. \label{eq:VectorRelation1}	
\end{empheq}
This formula for second-order elasticity tensors is completely analogous to  \eqref{eq:Relation}, which was obtained for 4$^{th}$ order tensors and allows to pass from micro to macro coefficients just by specifying the special forms of the $6\times6$ matrices $\Ct,\Cth,\Cte$. Using algebraic arguments analogous to those for the $4^{th}$ order tensors case, we obtain the inverse relation:
\begin{empheq}[box=\widefbox]{align}\Cte=\Cth \x \left(\Cth-\Ct\right)^{-1} \x \Ct\,. \label{eq:VectorRelation2}
\end{empheq}
These expressions may be of use when the elastic properties $\Cth$ of a unit elementary cell of the considered metamaterial and the macroscopic properties $\Ct$ of the metamaterial considered as a macroscopic block are known. Therefore, the elastic coupling tensor $\Cte$ is easily computable and is, in fact uniquely determined.
In the following subsections, we will particularize equations \eqref{eq:VectorRelation1} and \eqref{eq:VectorRelation2} to specific symmetries, thus dealing with isotropic, cubic, orthotropic an generally anisotropic materials, as intended in our relaxed micromorphic framework. For deriving such particular cases, we make the implicit assumption that $\Ce$, $\Ch$ and $\C$ have the same symmetries, which is indeed a sensible ansatz.

\subsubsection{The isotropic case}
{ In this subsection, we show how the fundamental formula \eqref{eq:VectorRelation1}	can be particularized to the isotropic case so retrieving the homogenization formulas for the Lamé parameters proposed in \cite{neff2007geometrically,neff2004material}.}

In the isotropic case and employing the Voigt notation, the constitutive elastic tensor has the following specific structure:
\begin{align} \Cte^{\mathrm{iso}}= & \left(\begin{array}{cccccc}2\me+\lle & \lle & \lle & 0 & 0 & 0\\\lle & 2\me+\lle & \lle & 0 & 0 & 0\\\lle & \lle & 2\me+\lle & 0 & 0 & 0\\0 & 0 & 0 & \me & 0 & 0\\0 & 0 & 0 & 0 & \me & 0\\0 & 0 & 0 & 0 & 0 & \me\end{array}\right),\end{align}
which, with the help of the bulk modulus $\ke=\frac{1}{3}(2\me+3\lle)$ can be expressed as:
\begin{align} \Cte^{\mathrm{iso}}= & \left(\begin{array}{cccccc}
\ke+4/3\,\me & \ke-2/3\,\me&  \ke-2/3\,\me & 0 & 0 & 0\\
 \ke-2/3\,\me & \ke+4/3\,\me&  \ke-2/3\,\me & 0 & 0 & 0\\
 \ke-2/3\,\me &  \ke-2/3\,\me & \ke+4/3\,\me & 0 & 0 & 0\\
0 & 0 & 0 & \me & 0 & 0\\
0 & 0 & 0 & 0 & \me & 0\\
0 & 0 & 0 & 0 & 0 & \me
\end{array}\right).
\end{align}
Analogously:
\begin{align}
\Cth^{\mathrm{iso}}= & \left(\begin{array}{cccccc}
\kh+4/3\,\mh & \kh-2/3\,\mh&  \kh-2/3\,\mh & 0 & 0 & 0\\
\kh-2/3\,\mh & \kh+4/3\,\mh&  \kh-2/3\,\mh & 0 & 0 & 0\\
\kh-2/3\,\mh & \kh-2/3\,\mh & \kh+4/3\,\mh & 0 & 0 & 0\\
0 & 0 & 0 & \mh & 0 & 0\\
0 & 0 & 0 & 0 & \mh & 0\\
0 & 0 & 0 & 0 & 0 & \mh
\end{array}\right).
\end{align}
Using the consistency condition in equation \eqref{eq:VectorRelation1} and simplifying, we can write:
\begin{align}
\Ct^{\mathrm{iso}}= & \left(\begin{array}{cccccc}
\km+4/3\,\mm & \km-2/3\,\mm&  \km-2/3\,\mm & 0 & 0 & 0\\
\km-2/3\,\mm & \km+4/3\,\mm&  \km-2/3\,\mm & 0 & 0 & 0\\
\km-2/3\,\mm & \km-2/3\,\mm & \km+4/3\,\mm & 0 & 0 & 0\\
0 & 0 & 0 & \mm & 0 & 0\\
0 & 0 & 0 & 0 & \mm & 0\\
0 & 0 & 0 & 0 & 0 & \mm
\end{array}\right),
\end{align}
where we set:
\begin{empheq}[box=\widefbox]{align}
\km=\frac{\ke\,\kh}{\ke+\kh}\,,\qquad \qquad \mm=\frac{\me\:\mh}{\me+\mh}  \label{eq:IsoA} \,.
\end{empheq}
The relation for $\km$ can also be expressed as a function of $\mm$ and $\lm$:
\begin{align} \left(2\mm+3\lm\right)=\frac{\left(2\mh+3\lh\right)\left(2\me+3\lle\right)}{\left(2\left(\me+\mh\right)+3\left(\lle+\lh\right)\right)} & . \label{eq:IsoB} \end{align}
Equations (\ref{eq:IsoA}) can also be inverted:
\vspace{-0.5cm}
\begin{empheq}[box=\widefbox]{align}
 \nonumber\\[+0.1cm]\ke=\frac{\km\:\kh}{\kh-\km}=\km\,\left(\kh-\km\right)^{-1}\,\kh\,,\nonumber\\[-0.2cm]\label{eq:IsotropicRel-1}\\
 \me=\frac{\mm\:\mh}{\mh-\mm}=\mm\,\left(\mh-\mm\right)^{-1}\,\mh\,. \nonumber\\[-0.2cm]\nonumber
\end{empheq}
 The first equation in \eqref{eq:IsotropicRel-1} can be analogously rewritten in terms of $\lle$ and $\me$ as:
  \begin{align} 
 \left(2\me+3\lle\right)&= \,\frac{\left(2\mm+3\lm\right)\left(2\mh+3\lh\right)}{\left(2\mh+3\lh\right)-\left(2\mm+3\lm\right)}\,.  \end{align}

\subsubsection{The cubic symmetry case}

{ In this subsection, we start showing the interest that the homogenization formula  \eqref{eq:VectorRelation1} may have in the case of simple anisotropies, as the cubic case. This formula for the cubic case will be applied in forthcoming works to show how it is fundamental for the mechanical characterization of real  metamaterials.}

In the cubic case, the constitutive tensors in Voigt-format have the following structure:

\begin{align} \Cte^{\mathrm{cub}}= & \left(\begin{array}{cccccc}2\me+\lle & \lle & \lle & 0 & 0 & 0\\\lle & 2\me+\lle & \lle & 0 & 0 & 0\\\lle & \lle & 2\me+\lle & 0 & 0 & 0\\0 & 0 & 0 & \me^{*} & 0 & 0\\0 & 0 & 0 & 0 & \me^{*} & 0\\0 & 0 & 0 & 0 & 0 & \me^{*}\end{array}\right)\,,\end{align}
which, using the bulk modulus $\ke=\frac{1}{3}(2\me+3\lle)$ can be rewritten as:
\begin{align} \Cte^{\mathrm{cub}}= & \left(\begin{array}{cccccc}
\ke+4/3\,\me & \ke-2/3\,\me&  \ke-2/3\,\me & 0 & 0 & 0\\
\ke-2/3\,\me & \ke+4/3\,\me&  \ke-2/3\,\me & 0 & 0 & 0\\
\ke-2/3\,\me & \ke-2/3\,\me & \ke+4/3\,\me & 0 & 0 & 0\\
0 & 0 & 0 & \me^{*} & 0 & 0\\
0 & 0 & 0 & 0 & \me^{*} & 0\\
0 & 0 & 0 & 0 & 0 & \me^{*}
\end{array}\right).
\end{align}
Analogously:
\begin{align}
\Cth^{\mathrm{cub}}= & \left(\begin{array}{cccccc}
\kh+4/3\,\mh & \kh-2/3\,\mh&  \kh-2/3\,\mh & 0 & 0 & 0\\
\kh-2/3\,\mh & \kh+4/3\,\mh&  \kh-2/3\,\mh & 0 & 0 & 0\\
\kh-2/3\,\mh & \kh-2/3\,\mh & \kh+4/3\,\mh & 0 & 0 & 0\\
0 & 0 & 0 & \mh^{*} & 0 & 0\\
0 & 0 & 0 & 0 & \mh^{*} & 0\\
0 & 0 & 0 & 0 & 0 & \mh^{*}
\end{array}\right).
\end{align}
Using the consistency condition in equation \eqref{eq:VectorRelation1}, we obtain:
\begin{align}
\Ct^{\mathrm{cub}}= & \left(\begin{array}{cccccc}
\km+4/3\,\mm & \km-2/3\,\mm&  \km-2/3\,\mm & 0 & 0 & 0\\
\km-2/3\,\mm & \km+4/3\,\mm&  \km-2/3\,\mm & 0 & 0 & 0\\
\km-2/3\,\mm & \km-2/3\,\mm & \km+4/3\,\mm & 0 & 0 & 0\\
0 & 0 & 0 & \mm^{*} & 0 & 0\\
0 & 0 & 0 & 0 & \mm^{*} & 0\\
0 & 0 & 0 & 0 & 0 & \mm^{*}
\end{array}\right).
\end{align}
where:
\begin{empheq}[box=\widefbox]{align}
\km=\frac{\ke\,\kh}{\ke+\kh}\,,\qquad\qquad \mm=\frac{\me\:\mh}{\me+\mh}, \qquad \qquad \mm^{*}=\frac{\me^{*}\:\mh^{*}}{\me^{*}+\mh^{*}} & .\label{eq:CubA}
\end{empheq}
The relation for $\km$ can also be expressed as a function of $\mm$ and $\lm$:
\begin{align} \left(2\mm+3\lm\right)=\frac{\left(2\mh+3\lh\right)\left(2\me+3\lle\right)}{\left(2\left(\me+\mh\right)+3\left(\lle+\lh\right)\right)} & . \label{eq:CubB} \end{align}
Equations (\ref{eq:CubA}) can also be inverted:
\vspace{-0.5cm}
\begin{empheq}[box=\widefbox]{align}
\nonumber\\[+0.1cm]\ke&=\frac{\km\:\kh}{\kh-\km}=\km\,\left(\kh-\km\right)^{-1}\,\kh\,,\nonumber\\[-0.2cm]\nonumber \\\label{eq:CubicRel-1}\me&=\frac{\mm\:\mh}{\mh-\mm}=\mm\,\left(\mh-\mm\right)^{-1}\,\mh,\\[-0.2cm]\nonumber\\\me^{*}&=\frac{\mm^{*}\:\mh^{*}}{\mh^{*}-\mm^{*}}=\mm^{*}\,\left(\mh^{*}-\mm^{*}\right)^{-1}\,\mh^{*}\,.\nonumber\\[-0.2cm]\nonumber
 \end{empheq}
 The first equation in \eqref{eq:CubicRel-1} can be analogously rewritten in terms of $\lle$ and $\me$ as:
 \begin{align} 
\left(2\me+3\lle\right)&= \,\frac{\left(2\mm+3\lm\right)\left(2\mh+3\lh\right)}{\left(2\mh+3\lh\right)-\left(2\mm+3\lm\right)}\,.  \end{align}

\subsubsection{The orthotropic case}

In the orthotropic case, the constitutive tensors have the following specific structure:
 \begin{align} \Cte^{\mathrm{orth}}= &\left(\begin{array}{cccccc} \left(\Cte\right)_{11} & \left(\Cte\right)_{12} & \left(\Cte\right)_{13} & 0 & 0 & 0\\ \left(\Cte\right)_{12} & \left(\Cte\right)_{22} & \left(\Cte\right)_{23} & 0 & 0 & 0\\ \left(\Cte\right)_{13} & \left(\Cte\right)_{23} & \left(\Cte\right)_{33} & 0 & 0 & 0\\ 0 & 0 & 0 & \left(\Cte\right)_{44} & 0 & 0\\ 0 & 0 & 0 & 0 & \left(\Cte\right)_{55} & 0\\ 0 & 0 & 0 & 0 & 0 & \left(\Cte\right)_{66} \end{array}\right)\,.
 \end{align}
Since it will be useful in the following, we define the sub-block $ \Cte^{\mathrm{a}}$ as:
 \begin{align}
 \Cte^{\mathrm{a}}= &\left(\begin{array}{cccccc} \left(\Cte\right)_{11} & \left(\Cte\right)_{12} & \left(\Cte\right)_{13} \\ \left(\Cte\right)_{12} & \left(\Cte\right)_{22} & \left(\Cte\right)_{23} \\ \left(\Cte\right)_{13} & \left(\Cte\right)_{23} & \left(\Cte\right)_{33}  \end{array}\right)\,
  \end{align}
 Analogously:
  \begin{align} \Cth^{\mathrm{orth}}= & \left(\begin{array}{cccccc} \left(\Cth\right)_{11} & \left(\Cth\right)_{12} & \left(\Cth\right)_{13} & 0 & 0 & 0\\  \left(\Cth\right)_{12} & \left(\Cth\right)_{22} & \left(\Cth\right)_{23}& 0 & 0 & 0\\  \left(\Cth\right)_{13} & \left(\Cth\right)_{23} & \left(\Cth\right)_{33}  & 0 & 0 & 0\\ 0 & 0 & 0 & \left(\Cth\right)_{44} & 0 & 0\\ 0 & 0 & 0 & 0 & \left(\Cth\right)_{55} & 0\\ 0 & 0 & 0 & 0 & 0 & \left(\Cth\right)_{66} \end{array}\right)\,,
  \end{align}
  and we define, for subsequent convenience, the sub-block $ \Cth^{\mathrm{a}}$ as:
  \begin{align} \Cth^{\mathrm{a}}= &\left(\begin{array}{cccccc} \left(\Cth\right)_{11} & \left(\Cth\right)_{12} & \left(\Cth\right)_{13} \\ \left(\Cth\right)_{12} & \left(\Cth\right)_{22} & \left(\Cth\right)_{23} \\ \left(\Cth\right)_{13} & \left(\Cth\right)_{23} & \left(\Cth\right)_{33}  \end{array}\right)\,. \end{align}
 Using the consistency condition in equation \eqref{eq:VectorRelation1}, we obtain:
   \begin{align} \Ct^{\mathrm{orth}}= & \left(\begin{array}{cccccc} \left(\Ct\right)_{11} & \left(\Ct\right)_{12} & \left(\Ct\right)_{13}   & 0 & 0 & 0\\  \left(\Ct\right)_{12} & \left(\Ct\right)_{22} & \left(\Ct\right)_{23} & 0 & 0 & 0\\  \left(\Ct\right)_{13} & \left(\Ct\right)_{23} & \left(\Ct\right)_{33}  & 0 & 0 & 0\\ 0 & 0 & 0 & \left(\Ct\right)_{44} & 0 & 0\\ 0 & 0 & 0 & 0 & \left(\Ct\right)_{55} & 0\\ 0 & 0 & 0 & 0 & 0 & \left(\Ct\right)_{66} \end{array}\right)\,,\end{align}
 where, considering $p=4,5,6$  without the sum over repeated indices, we have: 
 \begin{empheq}[box=\widefbox]{align} \Ct^{\mathrm{a}}=\Cte^{\mathrm{a}} \x \left(\Cte^{\mathrm{a}}+\Cth^{\mathrm{a}}\right)^{-1} \x\Cth^{\mathrm{a}}\,,\qquad\qquad \left(\Ct\right)_{pp}\text{=\,} & \frac{\left(\Cte\right)_{pp}\left(\Cth\right)_{pp}}{\,\left(\Cte+\Cth\right)_{pp}}\,.\label{eq:aniso}
 \end{empheq}
Here, we introduced the sub-block $\Ct^{\mathrm{a}}$:
  \begin{align} \Ct^{\mathrm{a}}= &\left(\begin{array}{cccccc} \left(\Ct\right)_{11} & \left(\Ct\right)_{12} & \left(\Ct\right)_{13} \\ \left(\Ct\right)_{12} & \left(\Ct\right)_{22} & \left(\Ct\right)_{23} \\ \left(\Ct\right)_{13} & \left(\Ct\right)_{23} & \left(\Ct\right)_{33}  \end{array}\right)\,.\nonumber \end{align}
The formulas in equation \eqref{eq:aniso} can also be inverted as:
\begin{empheq}[box=\widefbox]{align} \Cte^{\mathrm{a}}=\Ct^{\mathrm{a}} \x \left(\Cth^{\mathrm{a}}-\Ct^{\mathrm{a}}\right)^{-1}\x \Cth^{\mathrm{a}}\,,\qquad \left(\Cte\right)_{pp}=\,  \frac{\left(\Ct\right)_{pp}\left(\Cth\right)_{pp}}{\,\left(\Cth-\Ct\right)_{pp}}\,.
\end{empheq}

\subsection{The long wavelength limit - dynamic considerations}

The {governing} equations for the anisotropic relaxed micromorphic model {in the dynamical case} take the form:
\begin{align}
	\rho\, u_{,tt}=&\,\Div\left[\Ce \X \sym\left(\nablau-\p\right)+\Cc \X \skew\left(\nablau-\p\right)\right] ,\label{eq:EquilDyn} \\
	\nonumber
	\\
	\rho\,\widehat{L}_c^2 \,\Jfourth_{0} \X  \p_{,tt}=&\,\Ce \X \sym\left(\nablau-\p\right)+\Cc \X \skew\left(\nablau-\p\right)-\Ch \X\sym \p -\mLc\Curl \Curl \p .\nonumber 
\end{align}
Using the isotropy of inertia we can split \eqref{eq:EquilDyn}$_{2}$ into 3 coupled systems of equations:
\begin{align}
	\eta_{1}\, \rho\,\widehat{L}_c^2 \dev\,\sym \left[\p_{,tt}\right]=&\dev\,\sym\left[\Ce \X \sym\left(\nablau-\p\right)-\Ch \X\sym \p-\mLc\Curl \Curl \p \right]\,,\nonumber\\\nonumber\\
	\eta_{2}\, \rho\,\widehat{L}_c^2 \,\skew\left[\p_{,tt}\right]=&\Cc \X \skew\left(\nablau-\p\right)-\mLc \skew \Curl \Curl \p \,,\\\nonumber\\
	\eta_{3}\, \rho\,\widehat{L}_c^2 \tr\left[\p_{,tt}\right]=&\,\mathrm{tr}\,\left[\Ce \X \sym\left(\nablau-\p\right)-\Ch \X\sym \p -\mLc\Curl \Curl \p \right]\,.\nonumber
\end{align}
{This split of the inertia is essential for the description of real metamaterials in the dynamic regime (see \cite{madeo2016modeling}).}

The classical continuum theory is the \textbf{long wavelength limit}, corresponding to large length and time scales, and it predicts properties independent of specimen size. The long wave length limit is given by letting   $\widehat{L}_{c},L_{c}\rightarrow0$ simultaneously. In this case, the system \eqref{eq:EquilDyn} formally reduces to:
\begin{align}
	\rho\, u_{,tt}=&\,\Div\left[\Ce \X \sym\left(\nablau-\p\right)+\Cc \X \skew\left(\nablau-\p\right)\right] ,
	\label{eq:EquilDyn2}\\
	0=&\,\Ce \X \sym\left(\nablau-\p\right)+\Cc \X \skew\left(\nablau-\p\right) -\Ch \X\sym \p \,.\nonumber 
\end{align}
As in the static case, we may rewrite \eqref{eq:EquilDyn2} in the format of classical dynamic linear elasticity, yielding:
\begin{align}
\rho\, u_{,tt}=&\,\Div\left[\C \X\sym\nablau\right] , 
\end{align}
where, following \eqref{eq:Relation}, we obtain again:
\begin{align}
\C & =\Ch \X\left(\Ch+\Ce\right)^{-1} \X \, \Ce\,.
\end{align}

{ We have thus shown that the fundamental homogenization formula that we propose in this paper can be eventually obtained as a macroscopic limit in the statical case, or as a long wavelength limit in the dynamical case.}

\section{Non reduction for the standard Mindlin-Eringen model}
{In this section, we explicitly show that the considerations that allowed us to derive the macroscopic consistency conditions for the relaxed micromorphic model cannot be repeated for the classical Mindlin Eringen model which hence does not provide a transparent connection of the micro and meso elastic tensors to the macroscopic properties of the medium.}

The elastic energy of the general anisotropic micromorphic model in
the sense of Mindlin-Eringen can be represented as:
\begin{align}
 & W=\underbrace{\frac{1}{2}\langlenew\Coe \X \left(\nablau-\p\right),\left(\nablau-\p\right)\ranglenew}_{\mathrm{{\textstyle anisotropic\ elastic-energy}}}+\underbrace{\frac{1}{2}\langlenew\Ch \X\sym \p,\sym \p\ranglenew}_{\mathrm{{\textstyle micro-self-energy}}}\label{eq:EnerEringen-1} +\underbrace{\frac{\mLc}{2}\lVert  \nabla\hspace{-0.1cm} \p\rVert ^{2}}_{\textstyle \mathrm{curvature}}\,.
\end{align}
The same expression in index notation is:
\begin{align}
W= & \frac{1}{2}\left(\Coe\right)_{ijkl}\left(\nablau-\p\right)_{ij}\left(\nablau-\p\right)_{kl}+\frac{1}{2}\left(\Ch\right)_{ijkl}\left(\sym \p\right)_{ij}\left(\sym \p\right)_{kl} +\frac{\mLc}{2}\p_{ij,k} \p_{ij,k}.
\end{align}
Here, we have discarded $\E$ for simplicity. Note that the coupling of skew-symmetric terms is now also contained in $\Coe$ in some hidden way, instead of being explicitly present as in $\Cc$ and our relaxed model.
The static equilibrium equations are:
\begin{align}
\Div\left[\Coe \X \left(\nablau-\p\right)\right] & =0,
\label{eq:Equil-1}\\
\ -\Coe \X \left(\nablau-\p\right)+\Ch \X\sym \p+\mLc\,\Div\left[\nabla\hspace{-0.1cm} \p\right] & =0.\nonumber 
\end{align}
These can be equivalently written as:
\begin{align}
\left(\left(\Coe\right)_{ijkl}\left(\nablau-\p\right)_{kl}\right)_{,j} & =0,
\label{eq:Equil-1-1}\\
\ -\left(\Coe\right)_{ijkl}\left(\nablau-\p\right)_{kl}+\left(\Ch\right)_{ijkl}\left(\sym \p\right)_{kl}+\mLc\p_{ij,kk} & =0.\nonumber 
\end{align}
Here we can define the elastic (relative) stress in such a way that it depends bijectively on the non-symmetric elastic distortion $e=\nablau-\p$ since $\Coe$ is assumed to be uniformly positive definite:
\begin{align}
\sigma\left(\nablau,\p\right)=\,\Coe \X \left(\nablau-\p\right) \,,\qquad
\sigma_{ij}\left(\nablau,\p\right)=\,\left(\Coe\right)_{ijkl}\left(\nablau-\p\right)_{kl}  \,.\label{eq:SigmaMicro-2}
\end{align}
We can write in this model:
\begin{align}
\nablau-\p=\Coe^{-1}\hspace{-0.2cm}\X\, \sigma\,,
\end{align}
where $\Coe^{-1}$ is the Mindlin-Eringen elastic micromorphic compliance tensor.

In order to find the corresponding macroscopic tensor, we have to write the micromorphic elastic (relative) stress
as a function of only $\nablau$. 

Considering very large samples of the anisotropic structure
amounts to letting $L_{c}$, the characteristic length, tend to zero.
As a consequence of $L_{c}=0$, the second equilibrium equation in
(\ref{eq:Equil-1}) looses the $\Div\nabla\hspace{-0.1cm} \p$-term
and turns into an algebraic side-condition connecting $\p$
and $\nablau$ via:
\begin{align}
 & \Coe \X \left(\nablau-\p\right)=\Ch \X\sym \p\,.\label{eq:Equil2-1}
\end{align}
Or, again in index notation:
\begin{align}
 & \left(\Coe\right)_{ijkl}\left(\nablau-\p\right)_{kl}=\left(\Ch\right)_{ijkl}\left(\sym \p\right)_{kl}\,.\label{eq:Equil2-1-2}
\end{align}
From this equation we obtain:
\begin{align}
\Ch \X\sym \p=\, & \Coe \X \sym\left(\nablau-\p\right)+\Coe \X \skew\left(\nablau-\p\right) \nonumber\\
 =\,& \Coe \X \sym\nablau -\Coe \X \sym \p+\Coe \X \skew\left(\nablau-\p\right),
 \nonumber \\\nonumber\\
\left(\Coe+\Ch\right) \X \sym \p=\, & \Coe \X \sym\nablau +\Coe \X \skew\left(\nablau-\p\right)\nonumber \\
\nonumber \\
\sym \p= & \left(\Coe+\Ch\right)^{-1} \X \, \Coe \X \sym\nablau +\left(\Coe+\Ch\right)^{-1} \X \, \Coe \X \skew\left(\nablau-\p\right).  \label{eq:SymPErin}
\end{align}
In index notation this becomes:
\begin{align}
\left(\sym \p\right)_{ij}= & \left(\Coe+\Ch\right)_{ijkl}^{-1}\left(\Coe\right)_{klmn}\left(\sym\nablau \right)_{mn}+\left(\Coe+\Ch\right)_{ijkl}^{-1}\left(\Coe\right)_{klmn}\left(\skew\left(\nablau-\p\right)\right)_{mn}.
\end{align}
On the other hand, replacing \eqref{eq:Equil2-1} in \eqref{eq:Equil-1}$_{1}$
yields:
\begin{align}
\Div\left[\Ch \X\sym \p\right] & =0\,.
\end{align}
And again, by replacing this result in (\ref{eq:SymPErin}) we obtain:
\begin{align}
\Div\left[\Ch \X\left(\Coe+\Ch\right)^{-1} \X \, \Coe \X \sym\nablau +\Ch \X\left(\Coe+\Ch\right)^{-1} \X \, \Coe \X \skew\left(\nablau-\p\right)\right] & =0.
\end{align}
It is not possible to decouple this last equation due to the presence of the rotational coupling term $\skew\left(\nablau-\p\right)$. Therefore,
the only condition we can obtain is:
\begin{align}
\Ch \X\left(\Coe+\Ch\right)^{-1} \X \, \Coe \X \sym\nablau +\Ch \X\left(\Coe+\Ch\right)^{-1} \X \, \Coe \X \skew\left(\nablau-\p\right) & =\C \X\sym\nablau ,
\end{align}
or, in index notation:
\begin{align}
&\left(\Ch\right)_{klmn}\left(\Coe+\Ch\right)_{mnpq}^{-1}\left(\Coe\right)_{pqij}\left(\sym\nablau \right)_{ij}+\\&\qquad+\left(\Ch\right)_{klmn}\left(\Coe+\Ch\right)_{mnpq}^{-1}\left(\Coe\right)_{pqij}\left(\skew\left(\nablau-\p\right)\right)_{ij} =\left(\C\right)_{klij}\left(\sym\nablau \right)_{ij}.\nonumber 
\end{align}
This has to hold for any $\sym\nablau $.
Noting that $\C \X\sym\nablau \in\text{Sym}(3)$
and considering the symmetric part and the skew-symmetric part individually, we
have
\begin{align}
\begin{cases}
\ \sym\left\{ \Ch \X\left(\Coe+\Ch\right)^{-1} \X \, \Coe \X \sym\nablau +\Ch \X\left(\Coe+\Ch\right)^{-1} \X \, \Coe \X \skew\left(\nablau-\p\right)\right\} \\\qquad\qquad\qquad\qquad\qquad\qquad\qquad\qquad\qquad\qquad\qquad\qquad\qquad\qquad\qquad\qquad\qquad=\C \X\sym\nablau ,\\
\\
\skew\left\{ \Ch \X\left(\Coe+\Ch\right)^{-1} \X \, \Coe \X \sym\nablau +\Ch \X\left(\Coe+\Ch\right)^{-1} \X \, \Coe \X \skew\left(\nablau-\p\right)\right\} =0.
\end{cases}
\end{align}
Similarly, in index notation we obtain:
\begin{align}
\begin{cases}
\ \sym\left\{ \left(\Ch\right)_{klmn}\left(\Coe+\Ch\right)_{mnpq}^{-1}\left(\Coe\right)_{pqij}\left(\left(\sym\nablau \right)_{ij}+\left(\skew\left(\nablau-\p\right)\right)_{ij}\right)\right\} \ \\\hspace{10cm}=\left(\C\right)_{klij}\left(\sym\nablau \right)_{ij},\\
\\
\skew\left\{ \left(\Ch\right)_{klmn}\left(\Coe+\Ch\right)_{mnpq}^{-1}\left(\Coe\right)_{pqij}\left(\left(\sym\nablau \right)_{ij}+\left(\skew\left(\nablau-\p\right)\right)_{ij}\right)\right\} \ =0.\quad\ 
\end{cases}
\end{align}
A sufficient condition in order to obtain a decoupling of these equations
(sym and skew) is exactly the reduced anisotropic format put forward in our
relaxed model.

\section{The microscopic limit - static considerations}

There is another interesting limit behavior in our relaxed micromorphic model. We may consider, formally, to let  $L_{c}\rightarrow \infty$. Conceptually, this means a  ``zoom'' into the micro-structure. A scaling argument shows that this is tantamount to considering very small samples of the given multiscale material.

\subsection{The standard Mindlin-Eringen model}

Letting $L_{c}\rightarrow \infty$ and considering the curvature-term in the form $\mLc\lVert\nabla\hspace{-0.1cm} \p\rVert^2$ means that, in the limit, $\nabla\p\rightarrow0$ and $\p$ must be homogeneous: $\p(x)=\widehat{\hspace{-0.1cm}\p}$. This means that the micro-structure does not have the possibility to respond in any inhomogeneous way. The remaining minimization problem
\begin{align}
\int_{\Omega}&  \frac{1}{2}\langlenew\Coe \X \big(\nablau-\hspace{0.1cm}\widehat{\hspace{-0.1cm}\p}\big),\big(\nablau-\hspace{0.1cm}\widehat{\hspace{-0.1cm}\p}\big) \ranglenew +\frac{1}{2} \langlenew\Ch \X \sym\hspace{0.1cm}\widehat{\hspace{-0.1cm}\p},\sym\hspace{0.1cm}\widehat{\hspace{-0.1cm}\p}\ranglenew dx\rightarrow \min{\big(u,\hspace{0.1cm}\widehat{\hspace{-0.1cm}\p}\big)}\,,\qquad \hspace{0.1cm}\widehat{\hspace{-0.1cm}\p}\ \mathrm{is\ homogeneous}\,,\nonumber\\
 &\rightsquigarrow\begin{cases}
\Div \left[ \Coe \X \big(\nablau-\hspace{0.1cm}\widehat{\hspace{-0.1cm}\p}\big)\right]=0\,,\qquad u \vert_{\partial \Omega}= u_0\,,\\ \\
-\Coe\X\big(\nablau-\hspace{0.1cm}\widehat{\hspace{-0.1cm}\p}\big)+\Ch \X \sym\hspace{0.1cm}\widehat{\hspace{-0.1cm}\p}=0 \,, \qquad \hspace{0.1cm}\widehat{\hspace{-0.1cm}\p}\ \mathrm{is\ homogeneous,}
\end{cases}
\end{align}
 can be written, considering that $\hspace{0.1cm}\widehat{\hspace{-0.1cm}\p}$ is homogeneous, as
\begin{align}
\int_{\Omega} & \frac{1}{2}\langlenew\Coe \X \big(\nablau-\hspace{0.1cm}\widehat{\hspace{-0.1cm}\p}\big),\big(\nablau-\hspace{0.1cm}\widehat{\hspace{-0.1cm}\p}\big) \ranglenew  dx + \frac{\lvert\Omega\rvert}{2}  \langlenew\Ch \X \sym\hspace{0.1cm}\widehat{\hspace{-0.1cm}\p},\sym\hspace{0.1cm}\widehat{\hspace{-0.1cm}\p}\ranglenew\rightarrow\min(u,\hspace{0.1cm}\widehat{\hspace{-0.1cm}\p})\,,\nonumber\\
 &\rightsquigarrow\begin{cases}
\Div \left[ \Coe \X \nablau \right]=\underbrace{\Div \left[ \Coe \X \hspace{0.1cm}\widehat{\hspace{-0.1cm}\p}\right]}_{=0}=0\,, \qquad u \vert_{\partial \Omega}= u_0\,, \\
\Coe\X \hspace{0.1cm}\widehat{\hspace{-0.1cm}\p}+\Ch \X \sym\hspace{0.1cm}\widehat{\hspace{-0.1cm}\p}=\Coe\X\left[\frac{1}{\lvert\Omega\rvert}\int_{\Omega} \nablau dx \right] \,, \qquad \hspace{0.1cm}\widehat{\hspace{-0.1cm}\p}\ \mathrm{is\ homogeneous,} \label{eq:MinProb}
\end{cases}
\end{align} 
where $\lvert\Omega\rvert=\int_{\Omega}1\, dx$ denotes the measure of $\Omega$ and the last equation has been derived considering that the variation with respect to a homogeneous $\hspace{0.1cm}\widehat{\hspace{-0.1cm}\p}$ is:
\begin{align}
\int_{\Omega} & \langlenew\Coe \X \big(\nablau-\hspace{0.1cm}\widehat{\hspace{-0.1cm}\p}\big),-\delta\hspace{0.1cm}\widehat{\hspace{-0.1cm}\p} \ranglenew  dx + \lvert\Omega\rvert \langlenew\Ch \X \sym\hspace{0.1cm}\widehat{\hspace{-0.1cm}\p},\delta \hspace{0.1cm}\widehat{\hspace{-0.1cm}\p}\ranglenew \\\nonumber &= \int_{\Omega}  \langlenew\Coe \X \nablau ,-\delta\hspace{0.1cm}\widehat{\hspace{-0.1cm}\p} \ranglenew  dx +\int_{\Omega}  \langlenew\Coe \X \hspace{0.1cm}\widehat{\hspace{-0.1cm}\p},\delta\hspace{0.1cm}\widehat{\hspace{-0.1cm}\p} \ranglenew  dx  + \lvert\Omega\rvert \langlenew\Ch \X \sym\hspace{0.1cm}\widehat{\hspace{-0.1cm}\p},\delta \hspace{0.1cm}\widehat{\hspace{-0.1cm}\p}\ranglenew
\\\nonumber &= \langlenew \Coe \X\int_{\Omega}   \nablau dx ,-\delta\hspace{0.1cm}\widehat{\hspace{-0.1cm}\p} \ranglenew  +\lvert\Omega\rvert  \langlenew\Coe \X \hspace{0.1cm}\widehat{\hspace{-0.1cm}\p},\delta\hspace{0.1cm}\widehat{\hspace{-0.1cm}\p} \ranglenew   + \lvert\Omega\rvert \langlenew\Ch \X \sym\hspace{0.1cm}\widehat{\hspace{-0.1cm}\p},\delta \hspace{0.1cm}\widehat{\hspace{-0.1cm}\p}\ranglenew \\\nonumber &= \langlenew \Coe \X\int_{\Omega}   \nablau dx  +\lvert\Omega\rvert \left(  \Coe \X \hspace{0.1cm}\widehat{\hspace{-0.1cm}\p}  + \Ch \X \sym\hspace{0.1cm}\widehat{\hspace{-0.1cm}\p}\right),\delta \hspace{0.1cm}\widehat{\hspace{-0.1cm}\p}\ranglenew \,. 
\end{align}
The problem \eqref{eq:MinProb} has a unique solution in the displacement $u$, from which we determine $\hspace{0.1cm}\widehat{\hspace{-0.1cm}\p}$. Defining $\Cfourth:\R^{3\times3}\rightarrow\R^{3\times3}$ such that:
\begin{align}
\Cfourth \X \hspace{0.1cm}\widehat{\hspace{-0.1cm}\p}:=\Coe\X \hspace{0.1cm}\widehat{\hspace{-0.1cm}\p}+\Ch \X \sym\hspace{0.1cm}\widehat{\hspace{-0.1cm}\p}= \Coe\X \left[\frac{1}{\lvert\Omega\rvert} \int_{\Omega} \nablau dx \right]\,,
\end{align}
shows that $\Cfourth$ is invertible. Therefore, we obtain that the value of the homogeneous $\hspace{0.1cm}\widehat{\hspace{-0.1cm}\p}$ results as
\begin{align}
\hspace{0.1cm}\widehat{\hspace{-0.1cm}\p}:= \Cfourth^{-1} \hspace{-0.2cm} \X \ \Coe\X \left[\frac{1}{\lvert\Omega\rvert} \int_{\Omega} \nablau dx\right] \,.
\end{align}
In conclusion, the micro-distortion $\p$ is uniquely related to the average $\frac{1}{\lvert\Omega\rvert} \int_{\Omega} \nablau dx$ (over a representative unit cell). However, this relationship is not, in any way, transparent due to the unclear interaction of $\Cfourth$ and $\Coe$.

\subsection{The relaxed micromorphic model but with $\lVert\nabla\hspace{-0.1cm} \p\rVert^{2}$}
In this model variant, letting $L_{c}\rightarrow \infty$, generates again a response similar as before; $\p(x)=\widehat{\hspace{-0.1cm}\p}$ must be homogeneous and the remaining minimization problem
\begin{align}
\int_{\Omega}& \frac{1}{2} \langlenew\Ce \X \sym \big(\nablau-\hspace{0.1cm}\widehat{\hspace{-0.1cm}\p}\big),\sym \big(\nablau-\hspace{0.1cm}\widehat{\hspace{-0.1cm}\p}\big) \ranglenew +\frac{1}{2}  \langlenew\Ch \X \sym\hspace{0.1cm}\widehat{\hspace{-0.1cm}\p},\sym\hspace{0.1cm}\widehat{\hspace{-0.1cm}\p}\ranglenew  dx
\rightarrow \min{(u,\hspace{0.1cm}\widehat{\hspace{-0.1cm}\p})}\,,\nonumber\\
&\rightsquigarrow\begin{cases}
\Div \left[ \Ce \X \sym \big(\nablau-\hspace{0.1cm}\widehat{\hspace{-0.1cm}\p}\big)\right]=0\,,\qquad u \vert_{\partial \Omega}= u_0\,,\\ \\
-\Ce\X \sym \big(\nablau-\hspace{0.1cm}\widehat{\hspace{-0.1cm}\p}\big)+\Ch \X \sym\hspace{0.1cm}\widehat{\hspace{-0.1cm}\p}=0 \,, \qquad \hspace{0.1cm}\widehat{\hspace{-0.1cm}\p}\ \mathrm{is\ homogeneous,}
\end{cases}
\end{align}
can be written, since $\hspace{0.1cm}\widehat{\hspace{-0.1cm}\p}$ is homogeneous, as
\begin{align}
\int_{\Omega} & \frac{1}{2} \langlenew\Ce \X \sym  \big(\nablau-\hspace{0.1cm}\widehat{\hspace{-0.1cm}\p}\big),\sym \big(\nablau-\hspace{0.1cm}\widehat{\hspace{-0.1cm}\p}\big) \ranglenew  dx + \frac{\lvert\Omega\rvert}{2} \langlenew\Ch \X \sym\hspace{0.1cm}\widehat{\hspace{-0.1cm}\p},\sym\hspace{0.1cm}\widehat{\hspace{-0.1cm}\p}\ranglenew\rightarrow\min(u,\hspace{0.1cm}\widehat{\hspace{-0.1cm}\p})\,, \nonumber\\
&\rightsquigarrow\begin{cases}
\Div \left[ \Ce \X \sym \nablau \right]=\underbrace{\Div \left[ \Ce \X \sym \hspace{0.1cm}\widehat{\hspace{-0.1cm}\p}\right]}_{=0}=0\,, \qquad u \vert_{\partial \Omega}= u_0\,,\\
\left(\Ce+\Ch \right) \X \sym\hspace{0.1cm}\widehat{\hspace{-0.1cm}\p}=\Ce\X\left[\frac{1}{\lvert\Omega\rvert}\int_{\Omega} \sym \nablau dx \right]\,, \qquad \hspace{0.1cm}\widehat{\hspace{-0.1cm}\p}\ \mathrm{is\ homogeneous,} \label{eq:MinProb2}
\end{cases}
\end{align} 
where $\lvert\Omega\rvert=\int_{\Omega}1\,dx$
and the last equation has been derived by using the fact that the variation with respect to a homogeneous $\hspace{0.1cm}\widehat{\hspace{-0.1cm}\p}$ is:
\begin{align}
\int_{\Omega} & \langlenew\Ce \X \sym \big(\nablau-\hspace{0.1cm}\widehat{\hspace{-0.1cm}\p}\big),-\delta\hspace{0.1cm}\widehat{\hspace{-0.1cm}\p} \ranglenew  dx + \lvert\Omega\rvert \langlenew\Ch \X \sym\hspace{0.1cm}\widehat{\hspace{-0.1cm}\p},\delta \hspace{0.1cm}\widehat{\hspace{-0.1cm}\p}\ranglenew \\\nonumber &= \int_{\Omega}  \langlenew\Ce \X  \sym \nablau ,-\delta\hspace{0.1cm}\widehat{\hspace{-0.1cm}\p} \ranglenew  dx +\int_{\Omega}  \langlenew\Ce \X  \sym \hspace{0.1cm}\widehat{\hspace{-0.1cm}\p},\delta\hspace{0.1cm}\widehat{\hspace{-0.1cm}\p} \ranglenew  dx  + \lvert\Omega\rvert \langlenew\Ch \X \sym\hspace{0.1cm}\widehat{\hspace{-0.1cm}\p},\delta \hspace{0.1cm}\widehat{\hspace{-0.1cm}\p}\ranglenew
\\\nonumber &= \langlenew \Ce \X \int_{\Omega}   \sym \nablau dx ,-\delta\hspace{0.1cm}\widehat{\hspace{-0.1cm}\p} \ranglenew  +\lvert\Omega\rvert  \langlenew\Ce \X  \sym \hspace{0.1cm}\widehat{\hspace{-0.1cm}\p},\delta\hspace{0.1cm}\widehat{\hspace{-0.1cm}\p} \ranglenew   + \lvert\Omega\rvert \langlenew\Ch \X \sym\hspace{0.1cm}\widehat{\hspace{-0.1cm}\p},\delta \hspace{0.1cm}\widehat{\hspace{-0.1cm}\p}\ranglenew \\\nonumber &=  \langlenew \Ce \X\int_{\Omega}   \sym \nablau dx  +\lvert\Omega\rvert \left(  \Ce  + \Ch\right) \X \sym\hspace{0.1cm}\widehat{\hspace{-0.1cm}\p},\delta \hspace{0.1cm}\widehat{\hspace{-0.1cm}\p}\ranglenew \,. 
\end{align}
Since $\delta \hspace{0.1cm}\widehat{\hspace{-0.1cm}\p}$ is arbitrary, we obtain:
\begin{align}
\Ce \X\int_{\Omega}   \sym \nablau dx  +\lvert\Omega\rvert \left(  \Ce  + \Ch\right) \X \sym\hspace{0.1cm}\widehat{\hspace{-0.1cm}\p}=0\,.
\end{align} 
The problem \eqref{eq:MinProb2} has a unique solution in the displacement $u$, from which we determine $\sym\hspace{0.1cm}\widehat{\hspace{-0.1cm}\p}$:
\begin{align}
\sym \hspace{0.1cm}\widehat{\hspace{-0.1cm}\p}:= \left(\Ce+\Ch\right)^{-1} \X \,\Ce \X  \left[\frac{1}{\lvert\Omega\rvert} \int_{\Omega}\sym \nablau dx\right] \,,
\end{align}
from  which, since $\C=\Ch \X \left(\Ce+\Ch\right)^{-1} \X \,\Ce$,  we notice that:
\begin{align}
\Ch \X \sym \hspace{0.1cm}\widehat{\hspace{-0.1cm}\p}=&\,\C \X  \left[\frac{1}{\lvert\Omega\rvert} \int_{\Omega}\sym \nablau dx\right] \\
\iff\ \sym \hspace{0.1cm}\widehat{\hspace{-0.1cm}\p}=&\,\Ch^{-1} \X\, \C \X  \left[\frac{1}{\lvert\Omega\rvert} \int_{\Omega}\sym \nablau dx\right] \,.
\end{align}
Since $\Ch:\Sym(3)\rightarrow\Sym(3)$ and $\C:\Sym(3)\rightarrow\Sym(3)$ we also have that $\Ch^{-1}:\Sym(3)\rightarrow\Sym(3)$ and all together, $\Ch^{-1}\X\,\C:\Sym(3)\rightarrow\Sym(3)$. In this case, problem \eqref{eq:MinProb2} is formally equivalent to the classic elastic first gradient case when $\Csym=\Ce$. Since we used $\Cc \equiv 0$, the skew-symmetric part of $\p$ remains indeterminate.

\subsection{The relaxed micromorphic model with $\lVert\Curl \p \rVert^{2}$}
In our relaxed micromorphic model with the curvature depending only on $\Curl\p$, things turn out much differently. Letting $L_{c}\rightarrow\infty$ does not generate a homogeneous $\hspace{0.1cm}\widehat{\hspace{-0.1cm}\p}$; rather, it enforces that the micro-distortion  $\p$ must be compatible and therefore that there exists a function $\vartheta:\Omega\subset\R^{3}\rightarrow \R^{3}$ such that $\p(x)=\nabla \vartheta (x)$.

Therefore, the remaining minimization problem is:
\begin{align}
\int_{\Omega}& \frac{1}{2} \langlenew\Ce \X \sym \left(\nablau-\nabla \vartheta \right),\sym \left(\nablau-\nabla \vartheta\right) \ranglenew + \frac{1}{2} \langlenew\Ch \X \sym \nabla \vartheta,\sym \nabla \vartheta\ranglenew  dx\rightarrow \min{\left(u,\vartheta\right)}\,,\nonumber\\
&\rightsquigarrow\begin{cases}
\Div \left[ \Ce \X \sym \left(\nablau-\nabla \vartheta\right)\right]=0\,,\\ \\\Div \left[
-\Ce\X \sym \left(\nablau-\nabla \vartheta\right)+\Ch \X \sym\nabla \vartheta\right]=0\,.\label{eq:MinProb3}
\end{cases}
\end{align}
Leaving the boundary conditions for $\vartheta$ aside (i.e. no Dirichlet type boundary condition for $\p$) we immediately get a solution of \eqref{eq:MinProb3} by choosing $\nablau = \nabla \vartheta$ which gives:
\begin{align}
\int_{\Omega} \frac{1}{2} \langlenew\Ch \X \sym \nablau,\sym \nablau \ranglenew dx\rightarrow \min{\left(u\right)}\ \rightsquigarrow \ \Div \left[
\Ch \X \sym\nablau\right]=0\,.\label{eq:MinProb4}
\end{align}
Therefore, we get exactly the classical linear elastic response with the microscopic stiffness $\Ch$ for $L_{c}\rightarrow\infty$, as we should! 
 \begin{figure}
 \fcapside{\includegraphics[width=8.5cm]{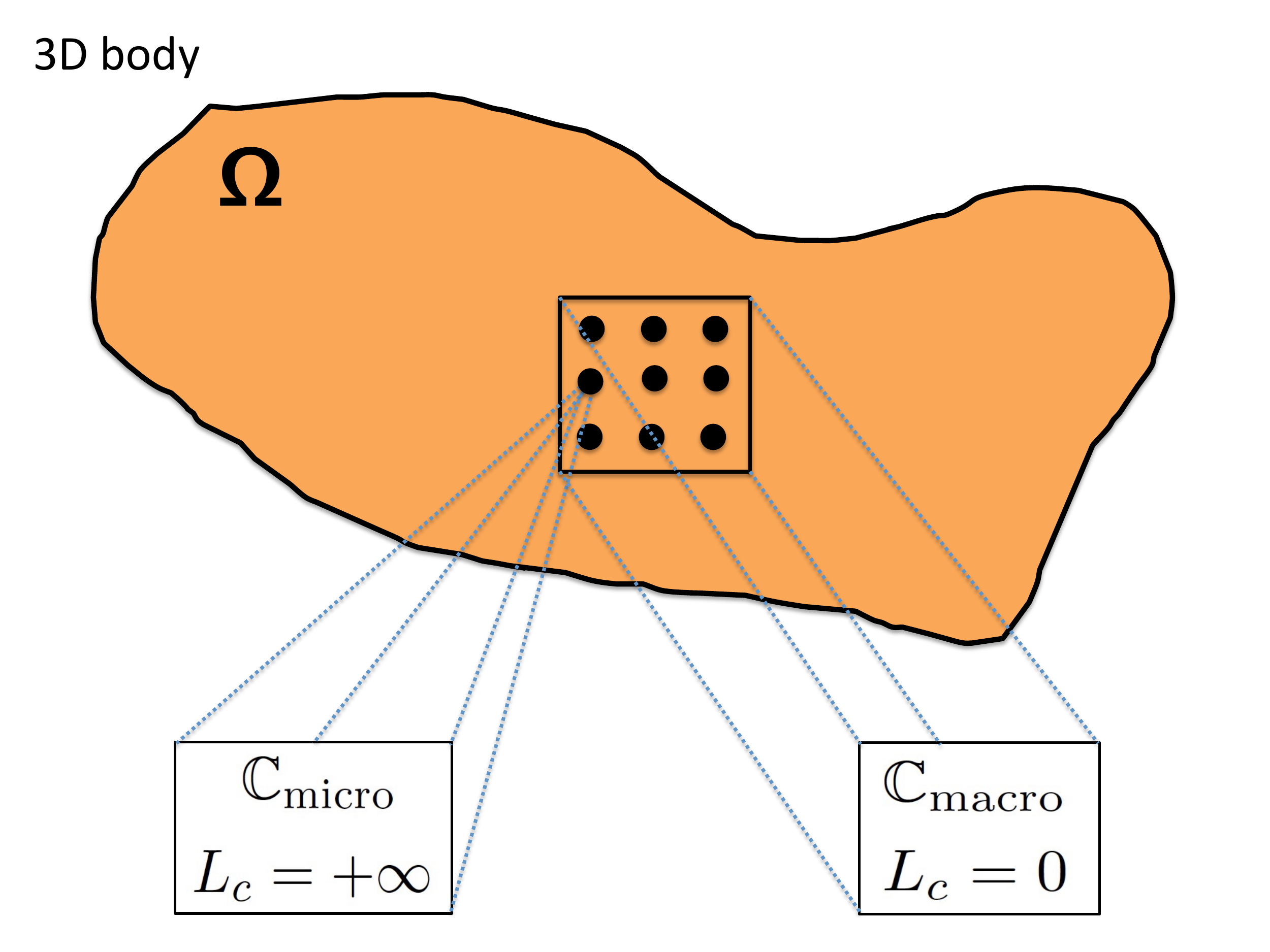}}{\caption{Transparent scale-separation between macroscopic Cauchy linear elastic response for $L_{c}=0$ with stiffness $\C$ and microscopic Cauchy linear elastic response for $L_{c}\rightarrow\infty$ with stiffness $\Ch$. The relaxed micromorphic model ``interpolates'' between $\Ch$ and $\C$ with parameter $L_{c}=0\rightsquigarrow\C$ and $L_{c}=\infty\rightsquigarrow\Ch$. This clearcut transparency is only obtained within our relaxed micromorphic approach.  Thereby, $L_{c}\rightarrow\infty$ solves only the micro-unit-cell problem while  $L_{c}=0$ can be interpreted either as making the body $\Omega$ arbitrary large while retaining the size of the unit-cell or keeping the dimensions of the body fixed while reducing the dimensions of the unit cell to zero. No condition on the rotational coupling tensor $\Cc$ is implied in either case.\label{fig:multiscale}}}
 \end{figure}


\section{Conclusion}

Gathering our new findings for the anisotropic relaxed micromorphic model together, we have obtained that for zero characteristic length scale $L_c=0$ (which corresponds to a long wavelength limit { or to a specimen of arbitrarily large size}), it is possible to identify both the symmetric and the skew-symmetric part of the micro-distortion $\p$ { as function of} to the gradient of the displacement $\nablau$:
\begin{align}
\left(\Ch+\Ce\right)\X\sym \p= & \,\Ce \X \sym\nablau \,,\\
\Cc\X \skew \left(\p\right)=&\,\Cc\X\skew \nablau \,. \nonumber
\end{align}
From this result we obtain that the experimentally observable macroscopic stiffness for an energy-equivalent linear elastic medium has the stiffness tensor:
\begin{align}
\C=\frac{1}{2} \mathcal{H}\left(\Ce,\Ch\right)=\left(\Ch^{-1}+\Ce^{-1}\right)^{-1} =\Ce\X\left(\Ch+\Ce\right)^{-1}\X \Ch.\label{eq:RelConc}
\end{align}
Here, $\mathcal{H}$ is the harmonic mean of the elastic (relative) stiffness tensor $\Ce$ and the microscopic stiffness tensor $\Ch$ of the relaxed micromorphic model. Inverting the expression \eqref{eq:RelConc} yields:
\begin{align}
\Ce=\Ch \X\left(\Ch-\C\right)^{-1} \X \, \C=\left(\C^{-1}-\Ch^{-1}\right)^{-1}  \,.\label{eq:RelConc2}
\end{align}
In \eqref{eq:RelConc2}, the tensor $\Ce$ is uniquely determined and positive definite, provided that $\Ch-\C$ is positive definite. No similar simple expression exists for the standard anisotropic Mindlin-Eringen model.

On the other hand, letting $L_{c}\rightarrow\infty$, we saw that our model tends to a classical linear elastic response with the microscopic stiffness $\Ch$. This results allows us to say that we have obtained a transparent scale-separation between the macroscopic linear response $\C$ for $L_{c}=0$ and the microscopic linear response $\Ch$ for $L_{c}\rightarrow\infty$. The intermediate cases can be interpreted as an  ``interpolation'' between the macro and the micro-behavior obtained for $L_{c}>0$.

We remark that the rotational coupling tensor $\Cc$ is in no way related to either the macroscopic or the microscopic measurable quantities, in sharp contrast to $\Ch$, $\C,\Ce$.

Furthermore, our presented model allows full use of the well-known Voigt-representation for classical elasticity tensors. Thus, we do not need to investigate the anisotropy classes based on 6$^{th}$-order tensors \cite{auffray2013algebraic}, neither for the local energy contribution nor for the curvature expression. This makes the presented framework by far more attractive, { due to the transparent comparison to classical linear, anisotropic elasticity.}

Our a priori novel macroscopic consistency condition \eqref{eq:RelConc} {drastically} reduces the burden of determining constitutive coefficients. {Indeed, the fundamental importance of formula \eqref{eq:RelConc} will be soon provided in a forthcoming paper in which a ``cubic'' band-gap metamaterial will be investigated. The macroscopic coefficients $\C$ will be determined on the basis of classical static tests on samples of the considered metamaterial. This will allow to drastically reduce the constitutive parameters to be determined. Such remaining parameters together with the micro-inertiae and, eventually the characteristic length $L_c$, will be determined on the basis of dynamical tests, following what done in \cite{madeo2016modeling} for the isotropic case.
	
	More particularly, the dispersion curves issued via the relaxed micromorphic model will be fitted on the experimental ones for some fixed directions of propagation of the traveling wave. Once the parameters will be calibrated, they will be validated by checking that the fitting on the dispersion curves remains reliable also on the other directions of propagation. This will provide the first evidence of the use of an enriched continuum model of the micromorphic type for the effective mechanical characterization of specific anisotropic metamaterials.}

\section{Acknowledgments}
The work of Ionel-Dumitrel Ghiba was supported by a grant of the Romanian National Authority for Scientific Research and Innovation, CNCS-UEFISCDI, project number PN-II-RU-TE-2014-4-0320. Angela Madeo wishes to thank INSA-Lyon for the funding of the BQR 2016 "Caractérisation mécanique inverse des métamatériaux: modélisation, identification expérimentale des paramètres et évolutions possibles", as well as the PEPS CNRS-INSIS.

\footnotesize
\let\stdsection\section
\def\section*#1{\stdsection{#1}}
\bibliography{library}
\bibliographystyle{plain}

\let\section\stdsection

\setcounter{section}{0}
\renewcommand\thesection{\Alph{section}}

\section{Appendices}
\subsection{Certain limiting cases of the relaxed micromorphic continuum \label{limiting}}

{In this section, we show} certain limiting cases of the anisotropic relaxed micromorphic continuum {model}. Since we assume $\Ch,\Ce$ to be positive definite and $\Cc$ positive semi-definite, 
there exist three positive constants $c^{+}_{\dev},c^{+}_{\tr},c^{+}_{e}>0$ and $c^{+}_{c}\geq0 $
such that:
\begin{align}
\langlenew \Ce \X \sym \left(\nablau-\p\right),\sym \left(\nablau-\p\right)\ranglenew_{\R^{3\times3}}&\geq c^{+}_{e}\lVert \sym \left(\nablau-\p\right) \rVert^{2}_{\R^{3\times3}}\,,\nonumber
\\\langlenew \Ch \X \sym \p,\sym \p\ranglenew_{\R^{3\times3}} &\geq c^{+}_{\dev}\lVert \dev \sym \p \rVert^{2}_{\R^{3\times3}}+c^{+}_{\tr}\left(\tr\left(\p\right)\right) ^{2}\,,\\\langlenew \Cc \X \skew \left(\nablau-\p\right),\skew \left(\nablau-\p\right)\ranglenew_{\R^{3\times3}}&\geq c^{+}_{c}\lVert \skew \left(\nablau-\p\right) \rVert^{2}_{\R^{3\times3}}
\nonumber.
\end{align}
Let us first consider:
\begin{align} \Ch\rightarrow\infty\,, \quad\Ce>0\,, 
\quad \Cc\geq 0\,,
\quad\p\in\R^{3\times3}\,,
\end{align}
which is the case if we assume $c^{+}_{\dev},c^{+}_{\tr}\rightarrow\infty$. In this case, the fact that the energy is bounded implies $\lVert\sym \p\rVert^2=0$ formally and, therefore, that $\p \in \so(3)$. This resulting model is equivalent to the \textbf{Cosserat model or micropolar model}. The appearance of only $\Curl\p$ in the curvature is consistent with the classical Cosserat or micropolar model, since for skew-symmetric $\p(x)=A(x) \in\so(3)$ it holds that $\Curl A$ is isomorphic to $\nabla A$, see \cite{neff2008curl}. On the other hand, we may consider:
\begin{align} 
\dev \sym\, \Ch\rightarrow\infty\,, \quad\Ce>0\,,
\quad \Cc\geq 0\,,
\quad\p\in\Sym(3)\,,
\end{align}
by which we mean to assume that $c^{+}_{\dev}\rightarrow\infty$ and  $\skew \p=0$. In this case, we obtain that $\lVert\dev \sym \p\rVert^2=0$ and, therefore, we can infer that $\p=\R\cdot \mathds{1}$. This model is called \textbf{micro-dilation theory} (see \cite{neff2014unifying}) and again, the presence of $\Curl\p$ is fully consistent with the general micro-dilation theory. One more case is:
\begin{align} 
\tr\, \Ch\rightarrow\infty\,, \quad\Ce>0\,,
\quad \Cc\geq 0\,,
\quad\p\in\R^{3\times3}\,,\label{eq:incomp}
\end{align}
where we assume that $c^{+}_{\tr}\rightarrow\infty$ and, therefore, $\tr \p=0$. In this case we obtain that $\p\in\mathfrak{sl}(3)$. This model is the \textbf{micro-incompressible micromorphic model}. Analogously, we may consider:
\begin{align} 
\dev \sym\, \Ch\rightarrow\infty\,, \quad\Ce>0\,,
\quad \Cc\geq 0\,,
\quad\p\in\R^{3\times3}\,,
\end{align}
where we assume that $c^{+}_{\dev}\rightarrow\infty$. In this case we obtain only that $\lVert\dev \sym \p\rVert^2=0$ and therefore that $\p=\R \cdot \mathds{1}+\so(3)$. This set of models is called \textbf{micro-stretch theory} (see \cite{neff2014unifying}).

Instead, if we just consider:
\begin{align} \Ch>0\,,  \quad\Ce>0\,, \quad \Cc\geq 0\,,
\quad\p\in\Sym(3)\,,
\end{align}
which means constraining $\p$ in such a way that $\skew \p=0$, then this resulting model is equivalent to Forest's  \textbf{microstrain model}, see \cite{forest2006nonlinear}. Finally, if we consider:
\begin{align} \Ch>0\,,\quad\Ce\rightarrow\infty\,, 
\quad \Cc=0\,,
\quad\p\in\Sym(3)\,.
\end{align}
by which we mean to assume that $c^{+}_{e}\rightarrow\infty$  and  $\skew \p=0$, we obtaine $\lVert\sym \left(\nablau-\p\right)\rVert^2=0$. Thus, it is possible to derive that $\sym\nablau=\sym\p=\p$. With this last property, we obtain that the curvature term reduces to $\langlenew\Lf_{	\mathrm{aniso}}\X\Curl\sym\nablau,\Curl\sym\nablau\ranglenew$. This resulting model is a variant of the indeterminate couple stress model, as treated in \cite{ghiba2016variant}.

It is not possible to suitably restrict the parameters of the relaxed micromorphic model in order to obtain a \textbf{full higher gradient elasticity} model, in sharp contrast to the standard Mindlin-Eringen model where $\me\rightarrow\infty,\mc\rightarrow\infty$ implies $\nablau=\p$ and $\lVert\nabla\hspace{-0.1cm}\p \rVert^2\rightarrow \lVert\nabla\nabla u\rVert^2$.

\subsection{One-dimensional standard Mindlin-Eringen model versus new relaxed micromorphic model \label{1D}} 

We let $u:[0,1]\rightarrow \R$ denote the displacement and $\widehat{p}:[0,1]\rightarrow \R$ the micro-distortion (we note that $u$ corresponds to the first component of the displacement and  $\widehat{p}$ corresponds to $\p_{11}$).

Considering a one-dimensional model, we can reduce the energy of the Mindlin-Eringen model to:
\begin{align}
\me |u'(t)-\widehat{p}(t)|^{2}+\mc|\underbrace{\skew(\cdot)}_{0}|^{2}+\mh\, |\widehat{p}(t)|^2+\frac{\mLc}{2}|\widehat{p}\,'(t)|^2\,.
\end{align}
Therefore, in a purely one-dimensional setting, the $\mc$-term does not appear. Furthermore, if $\me\rightarrow\infty$ formally, the energy reads:
\begin{align}
\mh \,|u'(t)|^{2}+\frac{\mLc}{2}\,|u''(t)|^2\,,\label{eq:2gr}
\end{align}
which is a second gradient elastic energy. The equilibrium equations read:
\begin{align}
2 \me \left(u'(t)-\widehat{p}(t)\right) \delta u'(t)&=0\,, \qquad \forall \, \delta u \in C^{\infty}_{0}([0,1],\R)\,, \nonumber\\ \\
\left[-2 \me \left(u'(t)-\widehat{p}(t)\right) +2\mh\, \widehat{p}  \right]\delta\widehat{p}(t)+\mLc\, \widehat{p}\,'\delta\widehat{p}\,'&=0\,.  \qquad \forall \,\delta \widehat{p} \in C^{\infty}_{0}([0,1],\R)\,, \nonumber
\end{align}
from which we obtain:
\begin{align}
\frac{d}{d\,x}\left[
2 \me \left(u'(t)-\widehat{p}(t)\right)\right]&=0\,, \qquad
-2 \me \left(u'(t)-\widehat{p}(t)\right) +2\mh\, \widehat{p} +\mLc \widehat{p}\,''=0\,. 
\end{align}
If we consider $L_{c}\rightarrow0$ we obtain: 
\begin{align}
\frac{\mathrm{d}}{\mathrm{d}\,x}\left[
2 \me \left(u'(t)-\widehat{p}(t)\right)\right]=0\,, \qquad-2 \me\, \left(u'(t)-\widehat{p}(t)\right) +2\mh\, \widehat{p}=0\,. 
\end{align}
This can be reduced to:
\begin{align}
\frac{\mathrm{d}}{\mathrm{d}\,x}\left[
2 \frac{\me \,\mh}{\me+\mh} u'(t)\right]=0\,, \qquad
\widehat{p}=\frac{\me}{\me+\mh}\, u'(t)\,\,.  
\end{align}
Therefore, this is equivalent to a classical elasticity model with energy:
\begin{align}
\mm\, |u'(t)|^2\,,\qquad \mathrm{with}\ 
\mm=\frac{\me\, \mh}{\me+\mh}\,.
\end{align}
Thus, in the one-dimensional setting, the Mindlin-Eringen format obeys our homogenization format as well.

For the relaxed micromorphic model we have instead:
\begin{align}
\me |u'(t)-\widehat{p}(t)|^{2}+\mc|\underbrace{\skew(\cdot)}_{=0}|^{2}+\mh\, |\widehat{p}(t)|^2+\frac{\mLc}{2}\Vert\underbrace{\Curl\left( \begin{array}{ccc}\widehat{p}& 0 & 0 \\ 0&0&0\\0&0&0\end{array}\right)}_{=0}\Vert^2\,.
\end{align}
Therefore, there are no terms with $L_{c}$ and the equilibrium equations read:
\begin{align}
\frac{\mathrm{d}}{\mathrm{d}\,x}\left[
2 \me \left(u'(t)-\widehat{p}(t)\right)\right]=0\,, \qquad
-2 \me \left(u'(t)-\widehat{p}(t)\right) +2\mh\, \widehat{p}=0\,. 
\end{align}
This is the the same format as the Mindlin-Eringen model with $L_{c}\rightarrow0$. 

Here, it must be noted that when $\me\rightarrow\infty$, we obtain formally only a first gradient elasticity model with energy:
\begin{align}
\mh \,|u'(t)|^2\,.
\end{align}
This is equivalent to a classical linear elasticity model with $\mm=\mh$, contrary to \eqref{eq:2gr}.

Here, one of the differences of the standard Mindlin-Eringen format, in comparison to the new relaxed formulation, clearly appears: the relaxed format does not reduce to a higher gradient elasticity model when specifying certain parameters.  

\subsection{Proof of equation \eqref{eq:Cu2} \label{Dem}}

By equation \eqref{eq:Cu} we have:
\begin{align}
\left(\Ce\right)_{ijkl}= &\underbrace{\mathfrak{M}_{\alpha ij} \left(\Cte\right)_{\alpha \beta}}_{A_{\beta ij}}\ \mathfrak{M}_{\beta kl}\,.\label{eq:Madeo0}
\end{align}
On the other hand, {using equation \eqref{eq:Cu}}, it can be seen that:
\begin{align}
A_{\beta i j}=\underbrace{\mathfrak{M}^{-1}_{q l \beta}\, \mathfrak{M}_{\alpha q l}}_{\widetilde{\delta}_{\alpha \beta}} A_{\alpha i j}=\mathfrak{M}^{-1}_{q l \beta} \left(\Ce\right)_{ij q l}\,,\label{eq:MadeoA}
\end{align}
and moreover, {formally introducing the tensor $A^{-1}$ such that $A_{\gamma m n}A^{-1}_{m n \beta}=\widetilde{\gamma}_{\beta \gamma}$, we also have}:
\begin{align}
\mathfrak{M}_{\beta hk}=\mathfrak{M}_{\gamma h k}\underbrace{ A_{\gamma m n}\, A^{-1}_{mn \beta}}_{\widetilde{\delta}_{\beta \gamma}}=\left(\Ce\right)_{mnhk} A^{-1}_{mn\beta}\,.\label{eq:MadeoB}
\end{align}
Using \eqref{eq:MadeoA} and \eqref{eq:MadeoB} in \eqref{eq:Madeo0} we get:
\begin{align}
\left(\Ce\right)_{ijhk}=A_{\beta ij}\, \mathfrak{M}_{\beta hk}=\mathfrak{M}^{-1}_{ql \beta} \left(\Ce\right)_{ijql} \left(\Ce\right)_{mnhk} A^{-1}_{mn \beta}=\left(\Ce\right)_{ijql}\, \mathfrak{M}^{-1}_{ql \beta}\, A^{-1}_{mn \beta} \left(\Ce\right)_{mnhk}
\,.
\end{align}
From this last expression, by comparing the first and the last equalities, we deduce:
\begin{align}
\mathfrak{M}^{-1}_{ql \beta}\, A^{-1}_{mn \beta} \left(\Ce\right)_{mnhk}=\mathds{1}_{qlhk}
\,.\label{eq:deltadelta}
\end{align}
Multiplying by $\left(\Ce^{-1}\right)_{hkrs}$, we get:
\begin{align}
\mathfrak{M}^{-1}_{ql \beta}\, A^{-1}_{mn \beta}\,\mathds{1}_{mnrs}
=\left(\Ce\right)^{-1}_{qlrs}\,. \label{eq:MadeoC}
\end{align}
In order to completely determine the fourth order tensor $\Ce^{-1}$ in terms of the second order {tensor $\Cte^{-1}$}, we need to write $A^{-1}$ explicitly. To this end we recall that, by definition, we have:
\begin{align}
A_{\beta ij}=\left(\Cte\right)_{\alpha \beta} \mathfrak{M}_{\alpha ij}\,.
\end{align} 
The following holds: \begin{align}
 A_{ ij \gamma}^{-1}= \mathfrak{M}^{-1}_{ij \delta} \left(\Cte\right)^{-1}_{\delta \gamma}\,.\label{eq:InvA}
 \end{align} 
Indeed, {using equation \eqref{eq:InvA} we can} compute:
\begin{align}
A_{\beta ij} \,A_{ ij \gamma}^{-1}=\left(\Cte\right)_{\alpha \beta} \mathfrak{M}_{\alpha ij}\,  \mathfrak{M}^{-1}_{ij \delta} \left(\Cte\right)^{-1}_{\delta \gamma}=\left(\Cte\right)_{\alpha \beta} \widetilde{\delta}_{\alpha \delta} \left(\Cte\right)^{-1}_{\delta \gamma}=\left(\Cte\right)_{\alpha \beta} \left(\Cte\right)^{-1}_{\alpha \gamma}=\widetilde{\delta}_{\beta \gamma}\,,
\end{align} 
where we used the symmetry of $\Cte$. This last chain of equalities guarantees that \eqref{eq:InvA} is actually the inverse of $A$. Then, replacing \eqref{eq:InvA} in \eqref{eq:MadeoC} we get:
\begin{align}
\left(\Ce\right)^{-1}_{qlrs}=\mathfrak{M}^{-1}_{ql \beta} \, \mathfrak{M}^{-1}_{mn \delta} \left(\Cte\right)^{-1}_{\delta \beta}\,\mathds{1}_{mnrs}
=\mathfrak{M}^{-1}_{ql \beta} \left(\Cte\right)^{-1}_{\delta \beta} \mathfrak{M}^{-1}_{rs \delta}\,.
\end{align}

\subsection{Some considerations about the anisotropic rotational coupling in the ``relaxed micromorphic model'' \label{Coupling}}

A method to reduce any given anisotropic rotational coupling to the isotropic case is, therefore, to simply project $\Ctc^{\mathrm{aniso}}$ to its isotropic part, given by the \textbf{arithmetic mean} of the eigenvalues of $\Ctc^{\mathrm{aniso}}$ (the Voigt bound):
\begin{align}
\mathrm{iso}_{\mathrm{arithm}}\left(\Ctc^{\mathrm{aniso}}\right):=\frac{1}{3} \mathrm{tr}\left(\Ctc^{\mathrm{aniso}}\right)\mathds{1}.\label{eq:CcMapp}
\end{align}
This defines a mapping $\mathrm{iso}_{\mathrm{arithm}}:\text{Sym}^{+}(3)\rightarrow \R^{+} \mathds{1}$.
We note, however, that applying \eqref{eq:CcMapp} has certain deficiencies, e.g. it is not stable under inversion:
\begin{align}
\mathrm{iso}_{\mathrm{arithm}}\left(\left(\Ctc^{\mathrm{aniso}}\right)^{-1}\right)\neq\left[\mathrm{iso}_{\mathrm{arithm}}\left(\left(\Ctc^{\mathrm{aniso}}\right)\right)\right]^{-1}.
\end{align}
It is possible, following the approach by Norris and Moakher \cite{moakher2006closest}, to obtain the closest isotropic tensor to $\Ctc^{\mathrm{aniso}}$ with respect to a geodesic structure on $\text{Sym}^{+}(3)$. This will define a nonlinear operator $\text{iso}_{\text{geod}}:\text{Sym}^{+}(3)\rightarrow \R^{+}\mathds{1}$ such that:
\begin{align}
\mathrm{iso}_{\text{geod}}\left(\left(\Ctc^{\mathrm{aniso}}\right)^{-1}\right)=\left[\mathrm{iso}_{\text{geod}}\left(\left(\Ctc^{\mathrm{aniso}}\right)\right)\right]^{-1}.
\end{align}
This will be exemplified in a different contribution. In the meantime, we may alternatively propose a mapping $\mathrm{iso}_{\text{log}}:\text{Sym}^{+}(3)\rightarrow \R^{+}\mathds{1}$ as:
\begin{align}
\mathrm{iso}_{\text{log}}\left(\Ctc^{\mathrm{aniso}}\right):=&\,e^{\frac{1}{3} \mathrm{tr}\left(\mathrm{log}\left(\Ctc^{\mathrm{aniso}}\right)\right)\mathds{1}}
=\,e^{\frac{1}{3} \mathrm{log}\left(\mathrm{det}\left(\Ctc^{\mathrm{aniso}}\right)\right)\mathds{1}}
=\,e^{\frac{1}{3} \mathrm{log}\left(\mathrm{det}\left(\Ctc^{\mathrm{aniso}}\right)\right)}\mathds{1}= \mathrm{det}\left(\Ctc^{\mathrm{aniso}}\right)^{\frac{1}{3}}\mathds{1}.
\end{align}
This is the \textbf{geometric mean} of the eigenvalues of $\Ctc^{\mathrm{aniso}}$. This mapping satisfies  
\begin{align}
\mathrm{iso}_{\text{log}}\left(\left(\Ctc^{\mathrm{aniso}}\right)^{-1}\right)=\left[\mathrm{iso}_{\text{log}}\left(\Ctc^{\mathrm{aniso}}\right)\right]^{-1}.
\end{align}
There is also another possibility. We define the harmonic isotropy projector by:
\begin{align}
\mathrm{iso}_{\mathrm{harm}}\left(\Ctc^{\mathrm{aniso}}\right):=\left[\mathrm{iso}_{\mathrm{arithm}}\left(\left(\Ctc^{\mathrm{aniso}}\right)^{-1}\right)\right]^{-1}.
\end{align}
This is the \textbf{harmonic mean} of the eigenvalues of $\Ctc^{\mathrm{aniso}}$ (the Reuss-bound \cite{bohlke2000minimum}). All introduced mappings satisfy the projection property:
\begin{align}
\mathrm{iso}_{\text{arithm}}\left(\gamma^{+}\mathds{1}\right)=\mathrm{iso}_{\text{geod}}\left(\gamma^{+}\mathds{1}\right)=\mathrm{iso}_{\text{log}}\left(\gamma^{+}\mathds{1}\right)=\mathrm{iso}_{\text{harm}}\left(\gamma^{+}\mathds{1}\right)=\gamma^{+}\mathds{1}.
\end{align}
Let us discuss the differences between $\mathrm{iso}_{\text{arithm}}$ and $\mathrm{iso}_{\text{log}}$. Consider a sequence of $\Ctc^{\mathrm{aniso},k}\rightarrow \Ctc^{\mathrm{aniso},\infty}$ for $k\rightarrow\infty$, where $\Ctc^{\mathrm{aniso},\infty}$ is \textbf{not} positive definite, i.e. some eigenvalue is zero (and $\mathrm{det}\left(\Ctc^{\mathrm{aniso},\infty}\right)=0$). Then:
\begin{align}
\mathrm{iso}_{\text{arithm}}\left(\Ctc^{\mathrm{aniso},k}\right)=\frac{1}{3} \mathrm{tr}\left(\Ctc^{\mathrm{aniso},k}\right)\mathds{1}\rightarrow\frac{1}{3} \mathrm{tr}\left(\Ctc^{\mathrm{aniso},\infty}\right) \mathds{1},
\end{align}
is positive definite. The mapping property is such that $\mathrm{iso}_{\text{arithm}}:\text{Sym}^{+}(3)\rightarrow \R^{+}\mathds{1}$. In contrast, we observe that:
\begin{align}
\mathrm{iso}_{\mathrm{log}}\left(\Ctc^{\mathrm{aniso},k}\right)=\left(\mathrm{det}\left(\Ctc^{\mathrm{aniso},k}\right)\right)^{\frac{1}{3}}\mathds{1}\rightarrow0_{\R^{3\times 3}}.
\end{align}
Therefore, $\mathrm{iso}_{\mathrm{log}}$ determines a zero isotropic coupling when eigenvalues of $\Ctc^{\mathrm{aniso}}$ vanish. For example, 
\begin{align}
\Ctc^{\mathrm{aniso}}=\left( \begin{matrix}
a_{1} & 0 & 0 \\0 & 0 & 0 \\ 0 & 0 & 0 
\end{matrix}\right),\qquad\qquad
\mathrm{iso}_{\text{arithm}}\left(\Ctc^{\mathrm{aniso}}\right)=\frac{a_{1}}{3},\qquad \qquad \mathrm{iso}_{\mathrm{log}}\left(\Ctc^{\mathrm{aniso}}\right)=0_{\R^{3\times 3}}.
\end{align}
At the present stage of understanding, however, we do not have extra arguments for using an anisotropic rotational coupling instead of an isotropic one. When possible, an isotropic rotational coupling given by the Cosserat couple modulus $\mc$ should be preferred.

\subsection{Properties of the resulting constitutive tensors \label{sec:Prop}}

\subsubsection{Symmetry}\label{sub:symm}

Let us first consider the direct relation:
\begin{align}
\C & =\Ch \X\left(\Ch+\Ce\right)^{-1} \X \, \Ce\, .
\end{align}
The constitutive tensor $\C$ is the result of a product of
the type:
\begin{align}
\C= &\, A\left(A+B\right)^{-1}B,
\end{align}
where $A$, $B$ and, as a consequence $\left(A+B\right)$ are symmetric.
In order to show the symmetry of $\C$ let us suppose that $\left(A+B\right)$
is invertible and write accordingly:
\begin{align}
\left(A+B\right) & \left(A+B\right)^{-1}B=B.
\end{align}
We can decompose the product by using the distributive
property of the matrix product with respect to the sum:
\begin{align}
& A\left(A+B\right)^{-1}B+B\left(A+B\right)^{-1}B=B.
\end{align}
Therefore:
\begin{align}
& A\left(A+B\right)^{-1}B=B-B\left(A+B\right)^{-1}B.
\end{align}
So we have that $\C=A\left(A+B\right)^{-1}B$ is the difference of
two symmetric matrices, since $B\left(A+B\right)^{-1}B$ is also symmetric.\footnote{We note again that the inverse of a positive definite tensor, like $A+B=\Ch+\Ce$ has the same symmetry group structure as $\Ch+\Ce$ itself. This can be easily shown by directly looking at the definition of groups.} For the inverse relation, we consider:
\begin{align}
\Ce & =\Ch \X\left(\Ch-\C\right)^{-1} \X \, \C\,.
\end{align}
Similarly, we can derive its symmetry (as long as $\left(\Ch-\C\right)_{kl}^{-1}$
exists):
\begin{align}
& A\left(A-B\right)^{-1}B=B+B\left(A-B\right)^{-1}B.
\end{align}

\subsubsection{Positive definiteness}

Let us now investigate the positive-definiteness of
\begin{align}
\C & =\Ch \X\left(\Ch+\Ce\right)^{-1} \X \, \Ce \,.
\end{align}
If we assume $\Ch$ and $\Ce$ to be positive definite,
it follows from the properties of positive definiteness, that their
sum as well as the inverse of the sum will be positive definite. Note
first that a product $AB$ of positive definite matrices $A$ and
$B$ has real, positive eigenvalues. This can be seen by considering the characteristic
equation:
\begin{align}
\mathrm{det}(A B-\lambda\mathds{1})=0\iff\mathrm{det}(A^{-1/2} [A B-\lambda\mathds{1}] A^{1/2})=0 & \iff\mathrm{det}(A^{1/2} B A^{1/2}-\lambda\mathds{1})=0\,.
\end{align}
Now, $A^{1/2} B A^{1/2}$ is positive definite since, setting $\eta:=A^{1/2}\xi$,
we have:
\begin{align}
\langlenew A^{1/2} B A^{1/2}\xi,\xi\ranglenew = & \langlenew B  A^{1/2}\xi,A^{1/2}\xi\ranglenew =\langlenew B\eta,\eta\ranglenew \geq\lambda_{\mathrm{min}}(B)\lVert \eta\rVert ^{2}=\lambda_{\mathrm{min}}(B)\lVert A^{1/2}\xi\rVert ^{2} \\
= & \,\lambda_{\mathrm{min}}(B)\langlenew A^{1/2}\xi,A^{1/2}\xi\ranglenew =\,\lambda_{\mathrm{min}}(B)\langlenew A\,\xi,\xi\ranglenew \geq\lambda_{\mathrm{min}}(B)\,\lambda_{\mathrm{min}}(A)\lVert \xi\rVert ^{2}\,. \nonumber
\end{align}
Therefore, the eigenvalues of $AB$ are real and positive. In general, however, the symmetry of the product $AB$ will be lost. In
our case, nonetheless, we proved in subsection \ref{sub:symm} that $\C$ is symmetric and,
therefore, positive definite.

For the inverse relationship, we consider:
\begin{align}
\Ce=\Ch \X\left(\Ch-\C\right)^{-1} \X \, \C \, .
\end{align}
In this case, in order to obtain the positive definiteness of $\Ce$
it is not enough to assume that $\Ch$ and $\C$
are positive definite. However, one sufficient condition to impose is
that $\Ch-\C$ is also positive definite. This
property can be thought of as a generalization of the condition found
in the isotropic case in which:
\begin{framed}
	\centering \textbf{smaller is stiffer}\\
	the macroscopic elastic response
	cannot be equal or stiffer than the microscopic response\\ \centering
	$\mh>\mm,\qquad \left(2\mh+3\lh\right)>\left(2\mm+3\lm\right)$.
\end{framed}

\end{document}